\documentclass[preprint2]{aastex}

\slugcomment{ApJ, in press}

\shorttitle{Variability of Be Stars}
\shortauthors{McSwain}

\begin{document}

\title{Variability of Be Stars in Southern Open Clusters}

\author{M. Virginia McSwain\altaffilmark{1,2}}
\affil{Department of Physics, Lehigh University, 16 Memorial Drive East, Bethlehem, PA 18015; mcswain@lehigh.edu}

\author{Wenjin Huang}
\affil{Department of Astronomy, University of Washington, Box 351580, Seattle, WA 98195-1580; wenjin@astro.washington.edu}

\author{Douglas R.\ Gies}
\affil{Center for High Angular Resolution Astronomy, Department of Physics and Astronomy, Georgia State University, P.O.\ Box 4106, Atlanta, GA 30302-4106; gies@chara.gsu.edu}

\altaffiltext{1}{Visiting Astronomer, Cerro Tololo Inter-American Observatory.  CTIO is operated by AURA, Inc.\ under contract to the National Science Foundation.}
\altaffiltext{2}{NSF Astronomy and Astrophysics Postdoctoral Fellow}


\begin{abstract}
We recently discovered a large number of highly active Be stars in the open cluster NGC 3766, making it an excellent location to study the formation mechanism of Be star disks.  To explore whether similar disk appearances and/or disappearances are common among the Be stars in other open clusters, we present here multiple epochs of H$\alpha$ spectroscopy for 296 stars in eight open clusters.  We identify 12 new transient Be stars and confirm 17 additional Be stars with relatively stable disks.  By comparing the H$\alpha$ equivalent widths to the photometric $y$--H$\alpha$ colors, we present a method to estimate the strength of the H$\alpha$ emission when spectroscopy is not available.  For a subset of 128 stars in four open clusters, we also use blue optical spectroscopy and available Str\"omgren photometry to measure their projected rotational velocities, effective temperatures, and polar surface gravities.  We combine our Be star detections from these four clusters to investigate physical differences between the transient Be stars, stable Be stars, and normal B-type stars with no line emission.  Both types of Be stars are faster rotating populations than normal B-type stars, and we find no significant physical differences between the transient and stable Be stars in our sample.  

\end{abstract}



\section{Introduction}

Classical Be stars are a class of B-type stars with circumstellar disks that cause Balmer and other line emission.  As a population, they rotate faster than their nonemission, B-type counterparts with a rotational velocity comparable to the critical velocity (\citealt{hunter2008}; \citealt{martayan2006}; \citealt{martayan2007}).  There are three possible reasons for the rapid rotation of Be stars: they may have been born as rapid rotators, spun up by binary mass transfer, or spun up during the main-sequence (MS) evolution of B stars.  The rotation is likely combined with weaker processes, such as nonradial pulsations (NRPs) or magnetic fields to move material from the stellar surface into the disk  \citep{porter2003}.  

We began studying the Be phenomenon in open clusters with a photometric and spectroscopic investigation of NGC 3766 (\citealt{mcswain2005a}; hereafter Paper 1).  We then applied our photometric technique to perform a survey of 55 southern open clusters to examine the Be star ages and dependence on main-sequence and post-main-sequence evolution (\citealt{mcswain2005b}; hereafter Paper 2).  With 52 definite Be star detections and 129 Be candidates identified in that survey, our results indicated that the fraction of Be stars in a cluster increases with age until 100 Myr.  Be stars are most common among the brightest, most massive B-type stars above the zero-age MS (ZAMS).  A spin-up phase at the terminal-age MS (TAMS) cannot produce the observed distribution of Be stars, but up to 73\% are consistent with spin-up by mass transfer in a close binary system.  Most of the remaining Be stars were likely rapid rotators at birth.  Other recent surveys of Be stars in Galactic and Magellanic Cloud open clusters have found similar results \citep{wisniewski2006, mathew2008}.

During a followup spectroscopic study of the cluster NGC 3766 (\citealt{mcswain2008a}; hereafter Paper 3), we discovered unprecedented activity among its Be stars.  Between 2003--2007, we observed disk appearances and/or near disappearances in 11 of its 16 Be stars.  The observed disk formation rates implied a very slow equatorial surface flow on average, but the surface flow can be an order of magnitude faster at times.  Our measurements of the changing disk sizes in NGC 3766 are consistent with the idea that transitory, nonradial pulsations contribute to the formation of these highly variable disks.

One of the primary motivations of this work was to confirm many of our Be candidates from Paper 2 with spectroscopic observations and improve the fraction of known Be stars in southern open clusters.  We also sought to discover additional Be stars that exhibit frequent changes in their disk states, transitioning between the Be and ``normal'' B-type spectra, as we observed among most of the Be stars in NGC 3766.   
Finally, spectroscopic observations allow us to probe the rotational velocities and other physical parameters that may distinguish Be stars from non-emitting B-type stars.  

Throughout this work, we adopt the term ``transient Be stars'' to describe any Be star observed in both the Be and normal B-type spectroscopic states.  Likewise, we use the term ``stable Be stars'' to describe Be stars whose disks are present in all observations, even though the strength of the disk emission may change dramatically.

Here we present multiple epochs of H$\alpha$ spectroscopy of eight open clusters, and we identify 12 previously unknown transient Be stars from this sample.  We also confirm 17 additional Be stars that have relatively stable disks.  For a subset of four open clusters, we also present blue optical spectroscopy and measurements of the effective temperature, $T_{\rm eff}$, surface gravity, $\log g$, and projected rotational velocity, $V \sin i$.  Combining these results with our previous measurements of members of NGC 3766, we compare the total samples of transient Be stars, stable Be stars, and normal B-type stars to explore physical differences in these three populations.  Other than the temporary disappearance of their disk emission, we find no physical differences between transient Be stars and the Be stars with relatively stable disks.



\section{Observations}

We obtained red optical spectra of 296 Be and B-type stars in 8 open clusters during observing runs in 2005 February, 2006 May, and 2007 May using the CTIO Blanco 4-m telescope with the Hydra multi-fiber spectrograph.  For a subset of 4 open clusters, we also obtained blue optical spectra during the 2006 and 2007 runs.  The spectrograph setups differed slightly between all these runs, and the details are summarized in Table \ref{spectroscopy}.  We used slit plates and/or order-sorting filters in some of our observing runs, and when either was not used, Table \ref{spectroscopy} contains a null value.

\placetable{spectroscopy}

We selected the targets for each run by giving highest priority to the known Be stars in the cluster.  We then selected other B-type stars in the cluster by ranking them according to their $y$--H$\alpha$ color to preferentially select any weak emission, Be star candidates from our photometry (Paper 2).  In Paper 3, we published results from 47 members of NGC 3766, and we here we expand that sample with red and blue spectra of 10 additional members observed in 2007 May.  Also in Paper 3, four Be stars did not have available blue spectra to measure their $V \sin i$, and we included them in the new target list.  For the other clusters, we used the same fiber assignments to achieve a target list as consistent as possible, but a broken fiber prohibited a few stars from being observed in later runs.  Also, the 2007 blue spectra suffered from much lower signal-to-noise ratios (S/N) due to the slit plate inserted in front of the fibers (possibly misaligned and blocking light), so several targets from this configuration were excluded from our measured sample.  

We generally began the Hydra observations by taking short exposures and then parking the fibers used for the brightest stars to avoid saturation in longer exposures of 1800 s.  Therefore we usually obtained up to 5--7 exposures of each star.  We also observed a HeNeAr comparison lamp source just before and after each set of cluster observations for wavelength calibrations.  

All of the spectra were zero corrected using standard routines in IRAF, and they were flat fielded, wavelength calibrated, and sky subtracted in IRAF using the \textit{dohydra} routine.  Of the 7 clusters observed, only NGC 3293 and IC 2581 had widespread nebulosity that was difficult to remove completely during the sky subtraction.  The effect was pronounced in the H$\alpha$ line in our red spectra, but the nebulosity presented much less of a problem in the H$\gamma$ and other Balmer lines in our blue spectra.  For each set of Hydra spectra, we transformed the observations to a common heliocentric wavelength grid, co-added them to achieve good S/N for each star, and rectified the total spectrum to a unit continuum.


\section{Be Star Variability and H$\alpha$ Equivalent Widths}

\begin{figure}
\includegraphics[angle=0,scale=0.8]{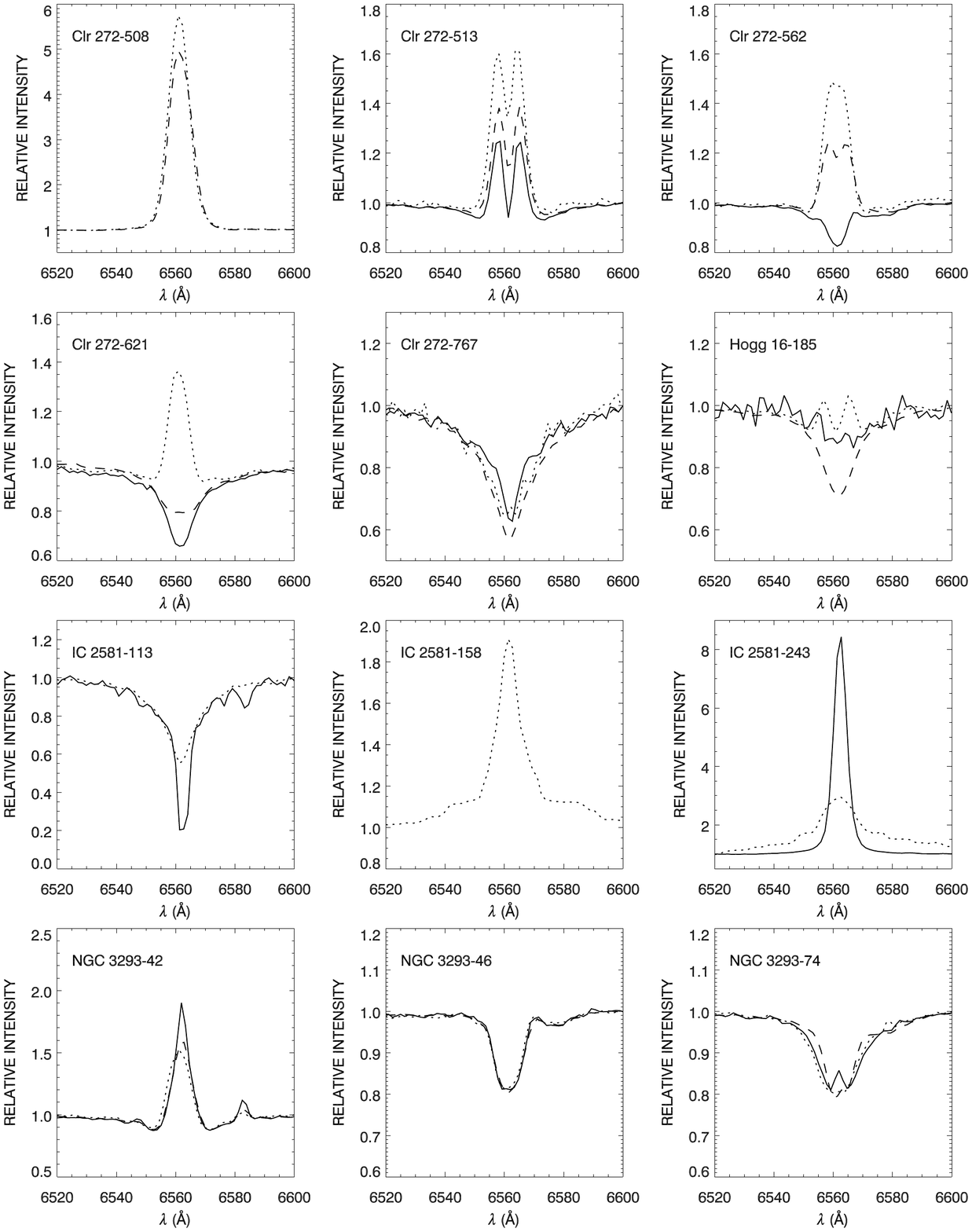}
\caption{H$\alpha$ profiles of Be stars, labeled by MG ID number.  Spectra from 2005, where available, are shown with dotted lines, spectra from 2006 with dashed lines, and spectra from 2007 with solid lines.  
\label{specvar1} }
\end{figure}

\begin{figure}
\includegraphics[angle=0,scale=0.8]{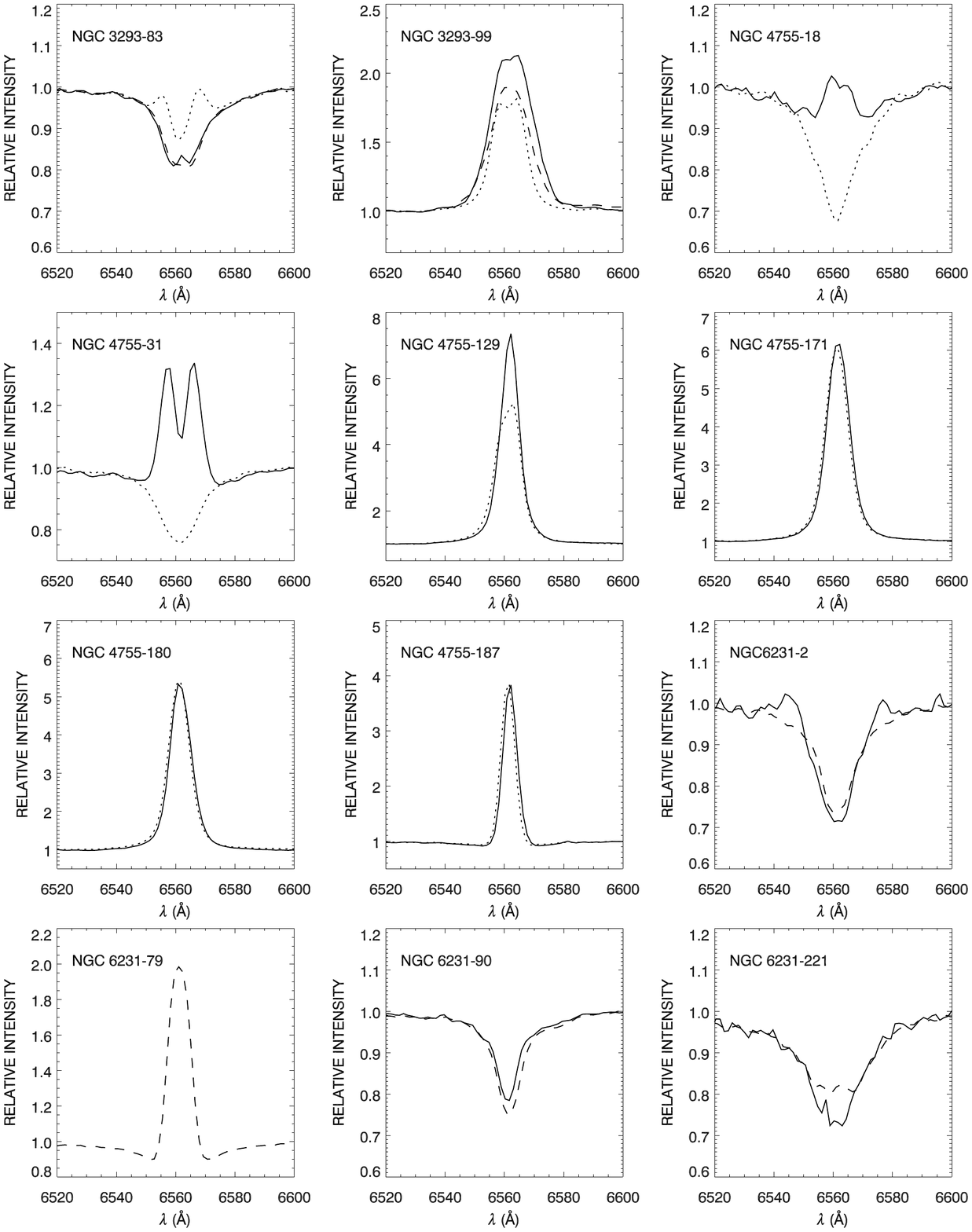}
\caption{H$\alpha$ profiles in the same format as Fig.\ \ref{specvar1}. 
\label{specvar2} }
\end{figure}

\begin{figure}
\includegraphics[angle=0,scale=0.8]{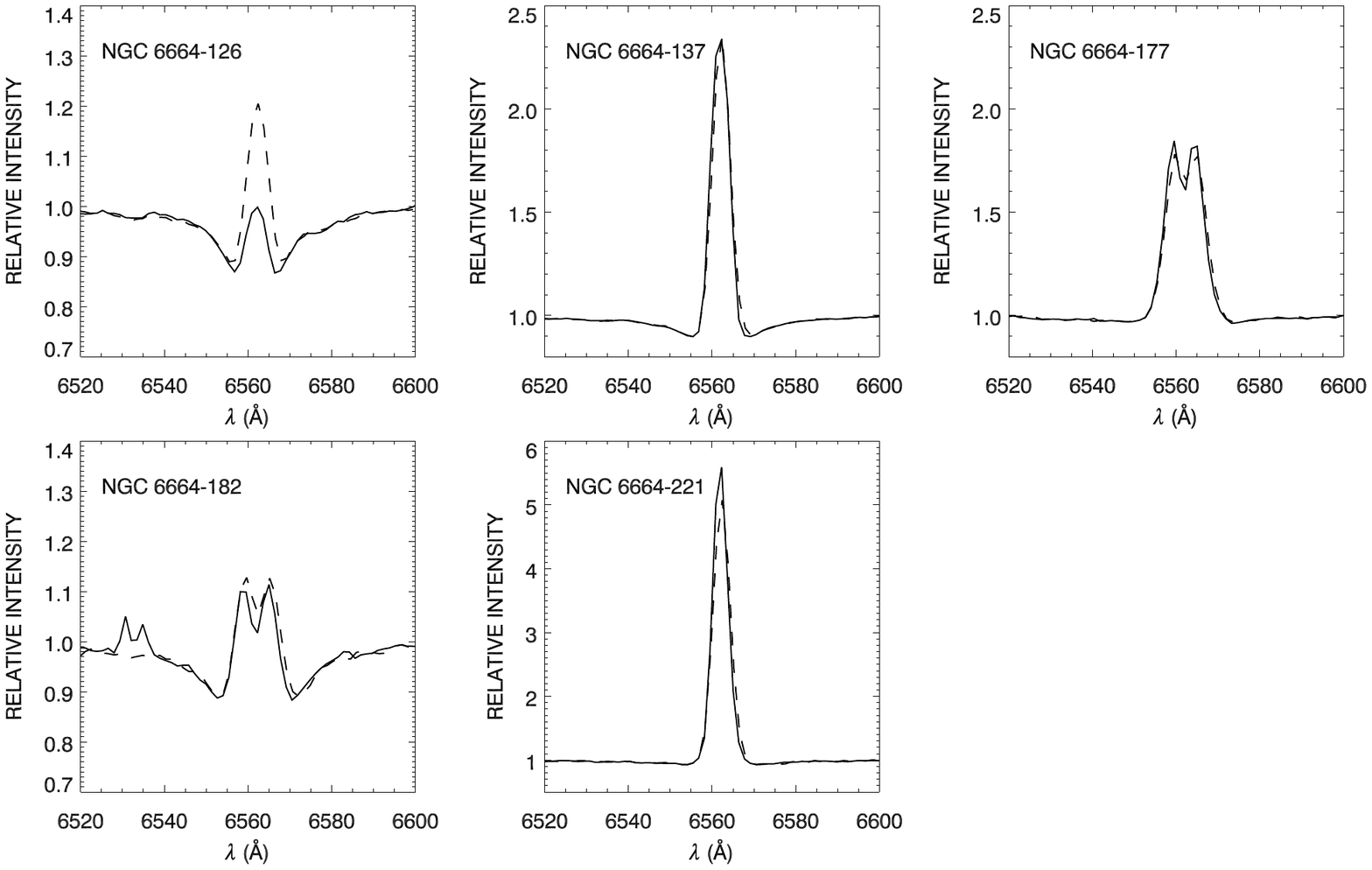}
\caption{H$\alpha$ profiles in the same format as Fig.\ \ref{specvar1}. 
\label{specvar3} }
\end{figure}

In Paper 3, we reported the discovery of 11 transient Be stars in NGC 3766 whose disks appeared and/or nearly completely disappeared during our 4 years of observations.  Only five of the observed Be stars had relatively consistent disks during the same period.  It turns out that NGC 3766 is not the only cluster containing such variable Be stars; during two or three epochs of red spectroscopy, we have discovered many new transient Be stars in nearly every cluster reported here.  The H$\alpha$ profiles of each Be star are shown in Figures \ref{specvar1}--\ref{specvar3}.

\placefigure{specvar1}
\placefigure{specvar2}
\placefigure{specvar3}

We measured the equivalent widths of the H$\alpha$ line profiles, $W_{H\alpha}$, by directly integrating across each rectified spectrum over the range 6535--6591 \AA~to include the broad wings expected for this strong line.  The error in $W_{H\alpha}$ is about 10\% due to noise in the continuum regions.  We use the convention that a pure absorption line has a positive equivalent width.  Many authors prefer to investigate Be emission strength by subtracting the photospheric absorption line profile and measuring $W_{H\alpha}$ from the net emission profile only.  We avoid that here for two reasons:  we would like to compare the raw $W_{H\alpha}$ for both emission line and non-emission line stars to their photometric colors, and we do not know an accurate photospheric line strength without knowing the effective temperature and gravity of the stars.  We measure these stellar properties below for a subset of our targets, but for consistency, we present only the raw $W_{H\alpha}$ for all stars.

The resulting $W_{H\alpha}$ are listed in Table \ref{eqwidths}, columns 2--4.  In column 5, we also identify the Be stars and candidates found in our photometry (Paper 2).  Finally in column 6, we list the Be star status based upon our red spectroscopic observations only.  While the photometric method is often unsuccessful identifying small or transient disks, it is usually very accurate when identifying strong emitters.  Only three stars were identified as Be in Paper 2 that are not confirmed here: Collinder 272-565, NGC 4755-43, NGC 6231-67.   We argue below that the first is likely a foreground star; the latter two may be Be transients that have not yet been confirmed spectroscopically.  

\placetable{eqwidths}

Several Be stars exhibit dramatic changes in their disk states, alternately showing strong disk emission and a near complete disappearance of the disk:  Collinder 272-562, Collinder 272-621, Hogg 16-185, NGC 3293-83, NGC 4755-18, and NGC 4755-31.  Normal B-type stars are not expected to have strong stellar winds with emission forming in the H$\alpha$ line, so any weak emission partially filling in the line or the wings is more likely due to a very small Be disk.  Several stars show line profile variations consistent with weak disks during one or more observations:  Collinder 272-767, NGC 3293-74, NGC 6231-2, NGC 6231-90, and NGC 6231-221.  Finally, the star IC 2581-113 appears to have a transient shell profile, indicating absorption through an edge-on disk.  We classify all 12 of these stars as Be transients.  Seven of these transient Be stars have available blue spectra, and combined with the 11 transient Be stars observed in NGC 3766, we have a large sample to compare the transient population with stable Be stars, discussed below.

An additional 17 Be stars exhibit line emission from their disks that are present in all of our observations, yet these ``stable'' disks vary in $W_{H\alpha}$ just as much or more than the transient disks.  In Paper 3, we found very similar mass gain and mass loss rates for the disks among all of the Be stars, suggesting similar origins for all of the disks.  The Be stars presented here have variability properties that are highly consistent with those observed in NGC 3766.

In total, we have identified 29 Be stars from our H$\alpha$ spectroscopy of Collinder 272/Hogg 16, IC 2581, NGC 3293, NGC 4755, NGC 6231, and NGC 6664, as well as 16 Be stars in NGC 3766 (Paper 3).  
Based on multiple epochs of H$\alpha$ spectroscopy for most of these 45 Be stars, we present lower limits of each cluster's transient Be star fraction in Table \ref{transient}.  Here, we calculate the percentage, $p$, of transient Be stars, $t$, compared to the total number of Be stars, $n$, using
\begin{equation}
p = \frac{t}{n}.
\end{equation}
Although the total populations of Be stars in each cluster are relatively small, the standard error associated with this fraction of transients is approximately given by the root of the variance for a binomial distribution,
\begin{equation}
\sigma_p \approx \sqrt{\frac{p(1-p)}{n}}.
\end{equation}
For NGC 6664, this standard error should be zero since $p=0$.  However, we find a more realistic error by allowing $p$ to vary until $\sigma_p \approx p$.  The Poisson errors associated with $t$ and $n$ are $\sqrt{t}$ and $\sqrt{n}$ respectively.  The clusters Collinder 272/Hogg 16, NGC 3766, and NGC 6231 have the highest rates of variable Be star disks.  

\placetable{transient}

\citet{meilland2006} discuss two mechanisms that may account for our observed transitions from the Be to B spectroscopic phases: discrete mass loss outbursts that form expanding ring-like structures around the star, and continuous mass loss that slowly decreases in magnitude and allows the disk to dissipate.  Based on our measurements of the changing disk masses in both stable and transient Be stars (Paper 3), we favor the first scenario.  In that work, we measured mass gain rates for the disks that appear to increase by an order of magnitude during short time intervals.  However, our spectra are not of sufficient resolving power to prove whether the emission lines evolve according to the ring scenario.  Higher resolution spectra will determine whether the peak separation in the emission lines remains constant while the high velocity wings disappear, as \citet{meilland2006} predict for such a discrete mass loss scenario.

\begin{figure}
\includegraphics[angle=90,scale=0.35]{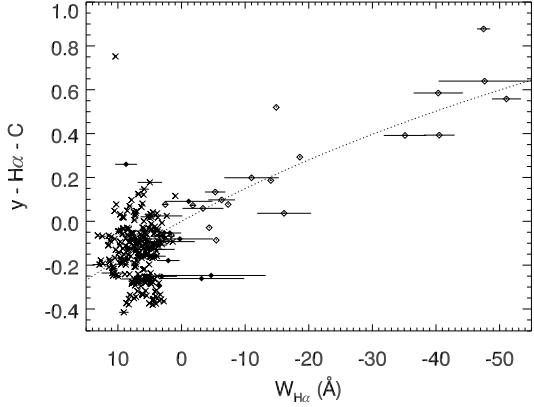}
\caption{The theoretical logarithmic relationship between $W_{\rm H\alpha}$ and $y$--H$\alpha$ color (\textit{dotted line}) is shown with the mean observed $W_{\rm H\alpha}$ and the adjusted $y$--H$\alpha$ color from Paper 2 for the transient Be stars (\textit{filled diamonds}), stable Be stars (\textit{open diamonds}), and non-emission B-type stars (\textit{crosses}).  
\label{eqw_color} }
\end{figure}

In Paper 1, we showed that in NGC 3766, the observed $W_{H\alpha}$ and $y-$H$\alpha$ colors have an approximately logarithmic relationship according to 
\begin{equation}
y - {\rm H}\alpha = C + 2.5 \log \frac{FWHM - W_{H\alpha}}{FWHM},
\end{equation}
where $FWHM$ is the range covered by the narrowband H$\alpha$ filter and $C$ is a constant.  Using our mean observed $W_{H\alpha}$ presented in Table \ref{eqwidths} and the $y-$H$\alpha$ colors from Paper 2, we have performed a similar fit for each cluster presented here.  The resulting fitting constants $C$ for each cluster are summarized in Table \ref{reddening}, and the adjusted colors $y-$H$\alpha - C$ for all stars with measured $W_{H\alpha}$ are plotted in Figure \ref{eqw_color}.  
The observed ranges in $W_{H\alpha}$ are shown with horizontal error bars.  Since $W_{H\alpha}$ is highly variable even among the stable Be stars, most of the scatter in this plot is explained by the time separation between our photometric and spectroscopic observations.  Some additional scatter is expected since the disks likely vary in density, opacity, and inclination.  The outlying star with $W_{H\alpha} = 10$~\AA~ (Collinder 272-565) is either a transient Be star or a foreground star (more likely since transient Be stars usually have bluer $y-$H$\alpha$ colors due to their smaller, temporary disks).

\begin{figure}
\includegraphics[angle=90,scale=0.35]{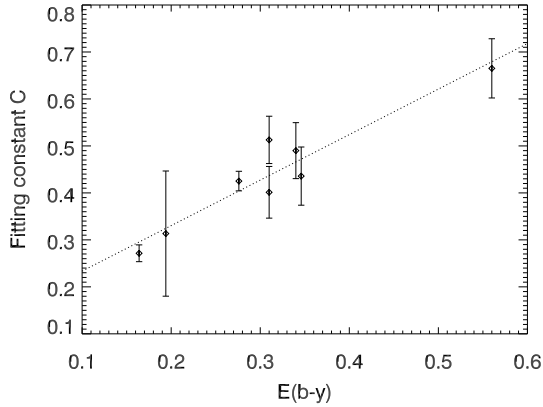}
\caption{The fitting constants, $C$, are plotted against the clusters' reddening values, $E(b-y)$.  The best fit linear relationship is also shown (\textit{dotted line}).  
\label{eqw_reddening} }
\end{figure}

\placetable{reddening}
\placefigure{eqw_color}
\placefigure{eqw_reddening}

Since reddening affects the continuum flux measured over the intermediate-band $y$ filter more than the narrow-band H$\alpha$ filter, the constant $C$ is dependent on the cluster's reddening, $E(b-y)$.  It will also depend upon the absolute scale of the H$\alpha$ magnitude.  We show in Figure \ref{eqw_reddening} that for our calibration of $y-$H$\alpha$, there is a direct relationship between $C$ and $E(b-y)$.  We find a linear fit
\begin{equation}
C = (0.97 \pm 0.12) E(b-y) + (0.14 \pm 0.04)
\end{equation}
between the two parameters.  Since spectroscopy is often unavailable in many observing programs, it is thus possible to estimate $W_{H\alpha}$ of a Be star from its photometric properties $y-$H$\alpha$ and $E(b-y)$.  A simple radiative transfer model such as that of \citet{grundstrom2006} may then be applied to investigate the physical size of the disk.  However, due to the large scatter in the observed relations we caution that absolute disk measurements are very imprecise and only differential changes in $y-$H$\alpha$ should be applied to estimate first-order changes in the disk's extent.  Since the H$\alpha$ filter system is not universally standard, different filters or photometric calibrations may produce a slightly different relationship between $C$ and $E(b-y)$ that must be determined before estimating $W_{H\alpha}$.


\section{Physical Parameters of Stars}

In Paper 3, we presented the physical parameters of 38 Be and B-type stars in NGC 3766 measured from blue optical spectra.  Here, we present measurements for additional members of NGC 3766 and three other clusters, for a total blue sample of 128 Be and B-type stars.  The blue sample includes only 17 of the transient Be stars and 12 of the stable Be stars, and we discuss the measurements of $V \sin i$, $T_{\rm eff}$, and $\log g_{\rm polar}$ below.

\subsection{$V \sin i$ Measurements}

Our technique for measuring $V \sin i$ for B-type stars is described in detail in Paper 3.  In short, we generated a grid of model spectra using SYNSPEC \citep{lanz2003} and Kurucz ATLAS9 atmospheric models \citep{kurucz1994} assuming plane-parallel geometry and local thermodynamic equilibrium (LTE).  We also adopted solar abundances and a microturbulent velocity of 2 km~s$^{-1}$ for all stars.  For each star, we made a preliminary estimate of its effective temperature and gravity, $T_{\rm eff}$ and $\log g$ respectively, by comparing the observed H$\gamma$, H$\delta$, and \ion{He}{1} line profiles to our grid of Kurucz spectral models.  To measure $V \sin i$, we compared the observed \ion{He}{1} $\lambda4388$ and \ion{He}{1} $\lambda4471$+\ion{Mg}{2} $\lambda4481$ line profiles to the model profiles convolved with a limb-darkened, rotational broadening function and a Gaussian instrumental broadening function.  We determined the best fit over a grid of values by minimizing the mean square of the deviations, rms$^2$.   The formal error, $\Delta V \sin i$, is the offset from the best-fit value that increases the rms$^2$ by $2.7 \, \rm rms^2$/$N$, where $N$ is the number of wavelength points within the fit region.  Our measured $V \sin i$ and $\Delta V \sin i$ are listed in columns 3--4 of Table \ref{params}.

\placetable{params}

In nearly all of the Be stars in our sample, the \ion{He}{1} lines do not exhibit obvious signs of emission.  These lines are, in general, much less susceptible to emission than the H$\gamma$ or H$\delta$ lines.  Nevertheless, we consider our $V \sin i$ measurements for Be stars to be lower limits since the \ion{He}{1} lines may contain some weak emission in Be stars, partially filling and narrowing their line profiles.  On the other hand, those clusters with strong nebular emission may have weak \ion{He}{1} emission that may cause these lines to appear more shallow, hence resulting in an overestimate of $V \sin i$.

We compared our measured $V \sin i$ to the measurements available for each cluster in the WEBDA database\footnote{The WEBDA database is maintained by J.-C. Mermilliod and is available online at http://obswww.unige.ch/webda/navigation.html.}.  The agreement is generally good, although we find a large scatter due to the systematic differences among authors.  For 18 stars, WEBDA lists $V \sin i$ measurements from multiple sources.  The differences between our measurements and WEBDA is comparable to the differences between the various sources.  We measured the standard deviation, $\sigma_{V \sin i}$, of the difference between our measurements and the literature to be 25.7 km~s$^{-1}$.  Stars with more than 3$\sigma_{V \sin i}$ difference in $V \sin i$ are probably unresolved double lined spectroscopic binaries (SB2s), and we find three new SB2 candidates in our sample: NGC 3293--27, NGC 6231--132, and NGC 6231--137.  We note that \citet{garcia2001} argue that NGC 6231--132 is an SB1 with a period of 8.9 d and that NGC 6231--137 is a $\beta$ Cep star and a probable SB1.  These and other spectroscopic binaries identified by blended lines in our spectra or by classification in the literature are identified in column 15 of Table \ref{params}.

\subsection{$T_{\rm eff}$ and $\log g$ Measurements of B stars}

We measured $T_{\rm eff}$ and $\log g$ for each B-type star using the technique described in Paper 3, and we summarize the procedure here.  For the coolest stars with $T_{\rm eff} < 15,000$ K, we used the ``virtual star'' method of \citet{huang2006b}, which assumes a spherically symmetric geometry and constant $T_{\rm eff}$ and $\log g$ across the stellar surface.    They generated detailed H$\gamma$ line profiles using line-blanketed, LTE Kurucz ATLAS9 and SYNSPEC codes.  Huang \& Gies show that the H$\gamma$ wing strength and line equivalent width can be used as starting parameters in a line profile fit to obtain unique values of $T_{\rm eff}$, $\log g$, and their corresponding errors.  

To measure the hotter B-type stars, we used the new TLUSTY BSTAR2006 grid of metal line-blanketed, non-LTE, plane-parallel, hydrostatic model spectra \citep{lanz2007}.  We used their models with solar metallicity and helium abundance and a microturbulent velocity of 2 km~s$^{-1}$.  We compared the H$\gamma$ line profile to the rotationally and instrumentally broadened model spectral line profiles at each value in the grid, minimizing rms$^2$ across the line region.  We then refined our measurements to a higher precision using a linear interpolation between the available line profiles in the grid.  Finally, we determined the errors, $\Delta T_{\rm eff}$ and $\Delta \log g$, from the values which produce a $\rm rms^2$ no more than $2.7 \, \rm rms^2$/$N$ greater than the minimum rms$^2$.  Our measurements of $T_{\rm eff}$ and $\log g$, with their corresponding errors, are listed in columns 5--8 of Table \ref{params}.  

For stars with $T_{\rm eff}$ near 15,000 K, we measured the parameters using both methods.  Since the method of Huang \& Gies can be applied to the hotter stars (although it does not include non-LTE effects), it is an appropriate tool to compare with TLUSTY in this temperature regime.  We find very good agreement from both methods, and we adopt the TLUSTY results here.  

A few stars in NGC 3293 were contaminated by nebulosity in the H$\gamma$ line, manifested as a narrow emission profile in the line core.  To measure $T_{\rm eff}$ and $\log g$ for these stars, we omitted the center 5--6 \AA~of the line profile in our fits and used only the broader wings.

Rapidly rotating B stars may be distorted into an oblate spheroidal shape, so the surface gravity at the equator can be much lower than at the poles, and this $\log g_{polar}$ is a better indicator of the evolutionary state of the star.  \citet{huang2006b} performed detailed spectroscopic modeling of such distorted rotating stars to determine a statistical correction factor for $\log g$, averaged over all possible $i$, for a variety of stellar models. We made a bilinear interpolation between their models to convert our measured $\log g$ to $\log g_{\rm polar}$ for a more accurate comparison between slow and rapid rotators.

For each B star, we also measured its mass, $M_\star$, and radius, $R_\star$, by interpolating between the evolutionary tracks for non-rotating stars from \citet{schaller1992}.  The errors $\Delta M_\star$ and $\Delta R_\star$ correspond to our measured $\Delta T_{\rm eff}$ and $\Delta \log g$.  For simplicity, we assume the polar radius of the star, $R_{\rm p}$ is equal to $R_\star$, and a rotationally distorted star has an equatorial radius $R_{\rm e} = 1.5 R_{\rm p}$ in the Roche approximation.  The resulting critical velocity is
\begin{equation}
V_{\rm crit} = \sqrt{\frac{G M_{\star}}{R_{\rm e}}}.
\end{equation}
Our results for $\log g_{\rm polar}$, $M_\star$, $\Delta M_\star$, $R_\star$, and $\Delta R_\star$, and $V_{\rm crit}$ are also listed in Table \ref{params}, columns 9--14.


\subsection{$T_{\rm eff}$ and $\log g$ Measurements of Be stars}

We measured $T_{\rm eff}$ and $\log g$ for the Be stars using available Str\"omgren $m_1$, $c_1$, and $\beta$ indices for our targets, available from the WEBDA database, as described in Paper 3.  To determine the dereddened indices, we adopted the reddening, $E(b-y)$, listed in Table \ref{reddening}.  We used the relation 
\begin{equation}
E(b-y) = 0.745 \; E(B-V)
\end{equation}
to relate $E(b-y)$ and $E(B-V)$ when the latter was given in the literature.  In Paper 3, we found that the temperature relation based on the dereddened Str\"omgren indices from \citet{balona1984} provides a fair agreement to our measured $T_{\rm eff}$ from the TLUSTY model fits for B stars with $T_{\rm eff} > 15,000$~K.  A small correction factor improves the agreement.  
We formed a sample of temperature calibrators using the B stars in NGC 3293, NGC 4755, and NGC 6231 with $T_{\rm eff} > 15,000$~K (measured from the TLUSTY models as described above) and available Str\"omgren photometry.  We also included eight calibrators with well-known $T_{\rm eff} > 15000$~K (measured from their absolute integrated stellar flux) from \citet{napiwotzki1993}.  We performed a linear fit to $T_{\rm eff}$ and $T_{\rm Balona}$ and found a very similar relationship to that found in Paper 3.  The photometry in WEBDA is from a number of different sources that often do not agree well, so the scatter in our temperature relation for these calibrators introduces an error of 1300 K, much higher than we found for NGC 3766 in Paper 3.  

To apply this temperature relation to the Be stars, we also need an accurate measurement of their $\beta$ magnitudes, which are unreliable due to their emission at this line.  As in Paper 3, we used the B star calibrators to find a linear fit to the observed ($c_0, \beta$) relation.  The mean scatter between $\beta_{\rm fit}$ and the measured $\beta$ is 0.017, which implies an additional error of 424 K in $T_{\rm eff}$ using the corrected $T_{\rm Balona}$ relation above.  

Finally, we measured $T_{\rm eff}$ for all Be stars in NGC 3293, NGC 4755, and NGC 6231 with Str\"omgren photometry available in the WEBDA database using the adopted $\beta_{\rm fit}$ and the corrected $T_{\rm Balona}$ as described above.  Adding the temperature errors from both fits in quadrature, we adopt a total error of 1400 K.  The results for the Be stars are listed in columns 5--6 of Table \ref{params}.  We found that all but one of the Be stars have $T_{\rm eff} > 15,000$~K, and the single exception has $T_{\rm eff} = 14,704$~K.  Therefore we did not need to extend the calibration sample to include stars with lower temperatures.  

The values of $\log g$ are more strongly dependent on $\beta$, so we were reluctant to use our $\beta_{\rm fit}$ with the Str\"omgren relation for $\log g$ given by \citet{balona1984}.  Instead, we used the calculated $T_{\rm eff}$ to determine the Be stars' bolometric corrections, BC, from \citet{lanz2007}, at first assuming $\log g=4.0$.  We calculated each stellar radius, $R_\star$, and luminosity, $L_\star$, using the measured $T_{\rm eff}$, BC, $V$ magnitude, distance modulus $(V-M_V)_0$, and $E(B-V)$.  The adopted $(V-M_V)_0$ are listed in Table \ref{reddening}.  We measured the stellar mass, $M_\star$, from the computed $T_{\rm eff}$ and $L_\star$ by interpolating between the evolutionary tracks of \citet{schaller1992}.  Finally, we obtained a preliminary value of $\log g$ from $M_\star$ and $R_\star$.  Since the BC is weakly dependent on $\log g$, we improved the BC from the initial estimate and iterated to compute the final $\log g$.  We adopt a formal error in $\log g$ computed from $\Delta T_{\rm eff}$ and the quoted errors in  $(V-M_V)_0$ and $E(B-V)$.  

We note that the evolutionary tracks of \citet{schaller1992} do not account for rotation, especially the very fast rotation observed among the Be stars.  Evolutionary tracks that include fast rotation reveal differences (sometimes very substantial) in the luminosity and MS lifetime of rapidly rotating stars (e.g., \citealt{ekstrom2008}).  However, the photometric magnitudes and colors of Be stars in our work are likely contaminated by enhanced bound-free and free-free emission from the circumstellar disks.  \citet{carciofi2006} showed that the $V$ magnitude of the Be star $\delta$ Sco increased by 0.7 magnitudes during a recent disk outburst.  However, they also found that the emission line strength and visual brightness in $\delta$ Sco are sometimes anticorrelated, possibly due to changes in the mass loss rate or disk geometry.  The uncertainty in $V$ combined with the large errors in our measured $T_{\rm eff}$ for the Be stars in this work negate any advantage in using evolutionary tracks for fast rotating stars.  Therefore we apply the Schaller et al.\ models throughout this work.

As we found in Paper 3, the agreement is very good between the Str\"omgren $\log g$ and $\log g_{\rm polar}$ for the nonemission B-type calibrators, so we do not perform any further correction to obtain $\log g_{\rm polar}$ for the Be stars.  However, several effects may cause the Be stars to appear slightly more evolved and artificially brightened.  Scattered light and continuum emission from the disk can increase the apparent brightness, an effect that is nearly impossible to correct without detailed knowledge of the disk geometry and orientation.  Rotational mixing in these rapid rotators can enrich the hydrogen-burning core and increase their luminosity during their evolution \citep{heger2000, meynet2000, ekstrom2008}.  On the other hand, the rapid rotation will also decrease the surface temperature at the stellar equator (the Von Zeipel effect; \citealt{vonzeipel1924}), and depending on the orientation, the flux may decrease.  We consider the values of $\log g_{\rm polar}$ to be lower limits, and our measurements of $R_\star$ upper limits, for the Be stars.  The final parameters for these Be stars are listed in Table \ref{params}.


\section{The Be and B Star Populations}

\subsection{Rotational Velocity Distributions}

\begin{figure}
\includegraphics[angle=90,scale=0.35]{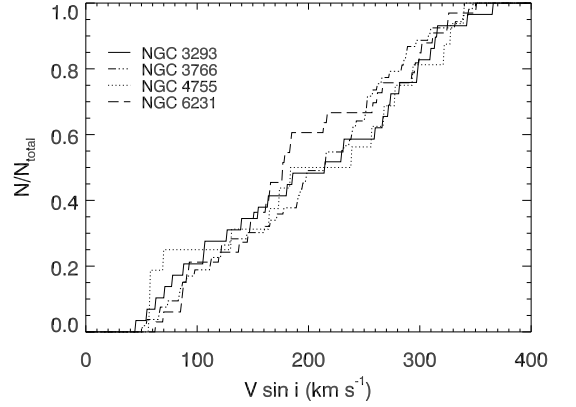}
\caption{Cumulative $V \sin i$ distribution for each cluster, including both Be and normal B-type stars.  
\label{vsini1} }
\end{figure}

The cumulative $V \sin i$ distribution for all Be and normal B-type stars in NGC 3293, NGC 3766, NGC 4755, and NGC 6231 are shown in Figure \ref{vsini1}, and we find that all four are very similar.  To test the null hypothesis that the distributions differ, we used the two-sided Kolmogorov-Smirnov (K-S) statistical test with all six permutations of  pairs of cluster data.  The K-S test indicates that no single cluster has a significantly different distribution of $V \sin i$.  
The known and candidate SB systems may have overestimated measurements of $V \sin i$ due to line blending, but they do not influence the relative distributions of the clusters' $V \sin i$.  We measure a mean $V \sin i = 184$~km~s$^{-1}$ with a standard deviation of 90~km~s$^{-1}$ for the normal B-type stars, including the SB systems.

\placefigure{vsini1}

A number of environmental factors may influence rotational velocities among cluster members: the ages, binary fraction, and magnetic fields. 
\citet{huang2006b, huang2008} find that the most massive B-type stars spin down with age as predicted by models due to their radial expansion and angular momentum lost by stellar winds.  Since the four clusters we observed have nearly identical ages (Paper 2), we are not surprised that their $V \sin i$ distributions are very similar.  
Binary systems are expected to spin down faster than single star systems due to tidal synchronization processes \citep{abt2002}, and \citet{huang2006b} found a sharp decline in $V \sin i$ for binaries in which the more massive component has $\log g_{\rm polar} \le 3.9$.  The cluster NGC 6231 has a very high binary fraction of at least 52\% among its B-type stars \citep{raboud1996} and at least 63\% among O-type stars \citep{sana2008}, so many of our targets are likely binaries that will lose angular momentum due to tidal interactions as they evolve.  However, most of our targets have $\log g_{\rm polar} > 4.0$ and do not appear to have reached synchronization.
Finally, \citet{huang2006b} found that intermediate and low mass B stars spin down faster than more massive B-type stars, possibly due to additional angular momentum losses such as magnetic fields.  Indeed, two magnetic B stars were recently detected in NGC 3766 \citep{mcswain2008b}.  
The similarity in the $V \sin i$ distributions in NGC 3293, NGC 3766, NGC 4755, and NGC 6231 suggests that the environments are very similar or that the clusters are too young for the various environmental factors to produce observable differences in rotation.

\begin{figure}
\includegraphics[angle=90,scale=0.35]{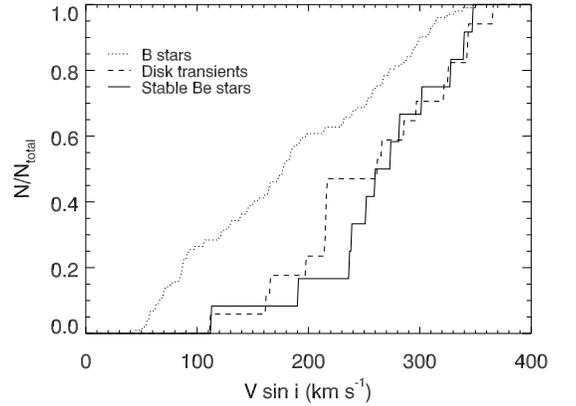}
\caption{Cumulative $V \sin i$ distribution for Be stars with stable disks and those with transient disks.  The distribution for normal B-type stars is also shown for reference.
\label{vsini2} }
\end{figure}

Since we find no systematic differences among the cluster $V \sin i$ distributions, we can explore trends in $V \sin i$ for the combined Be star populations.  We have measured $V \sin i$ for 12 Be stars with relatively steady-state disks (emission present in each of our observations) and 17 Be stars with significant changes in their disk emission state (either outbursts or disk fading events).  The mean $V \sin i$ and its standard deviation for the stable (transient) Be stars is $263 \pm 66$~km~s$^{-1}$ ($253 \pm 74$~km~s$^{-1}$).  Figure \ref{vsini2} shows the cumulative distribution of $V \sin i$ for the Be stars with stable disks and transient disks compared to the normal B-type cluster population.  The K-S test indicates a relatively high probability (46\%) that the transient and stable Be stars have the same $V \sin i$ distribution, so the differences in their distributions are not significant.  For the combined Be star sample, the K-S test indicates only a 0.03\% probability that their $V \sin i$ are drawn from the same distribution as the normal B-type stars.  Both the stable and transient Be stars are clearly populations of rapid rotators compared to the normal B-type stars in the sample, in excellent agreement with several other recent studies of Be rotation (\citealt{hunter2008}; \citealt{martayan2006}; \citealt{martayan2007}; Paper 3).  

\placefigure{vsini2}

\subsection{$T_{\rm eff}$ and $\log g$ Distributions}

\begin{figure}
\includegraphics[angle=90,scale=0.35]{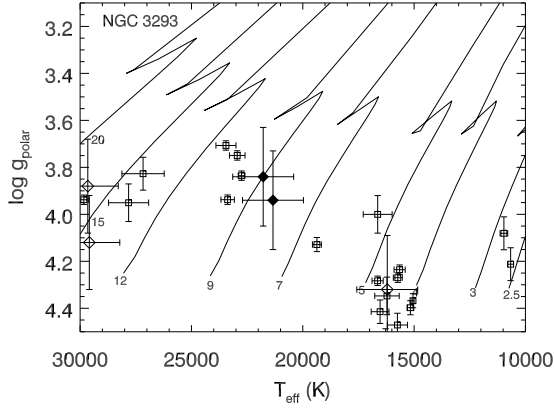}
\caption{$T_{\rm eff}$ and $\log g$ are plotted for the cluster NGC 3293 with the evolutionary tracks of Schaller et al.\ (1992; \textit{solid lines}).  The ZAMS mass of each evolutionary track is labeled along the bottom.  Normal B-type stars are shown as open squares, stable Be stars as open diamonds, and Be transients as filled diamonds. 
\label{tlogg3293} }
\end{figure}

\begin{figure}
\includegraphics[angle=90,scale=0.35]{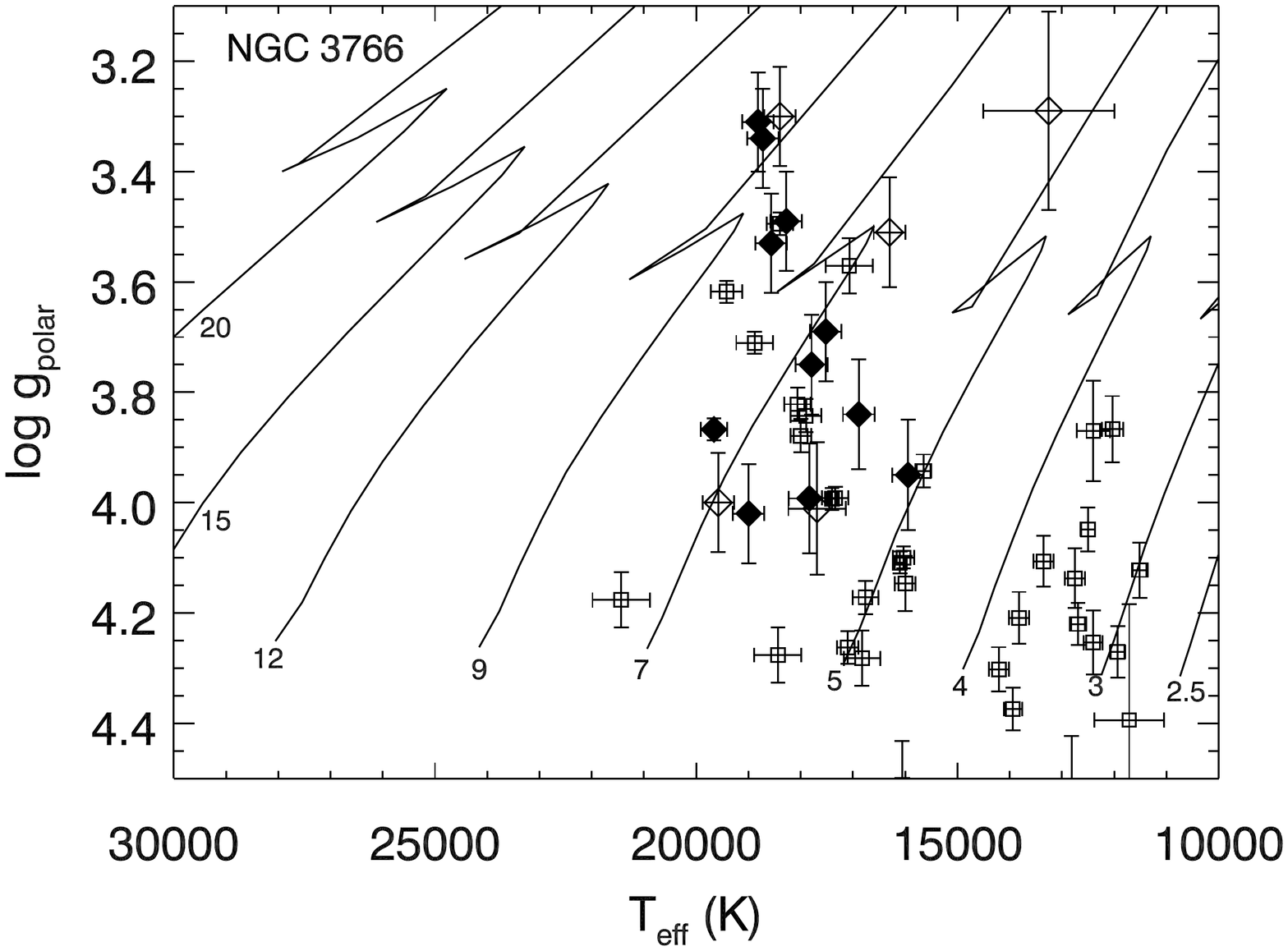}
\caption{$T_{\rm eff}$ and $\log g$ are plotted for the cluster NGC 3766 in the same format as Figure \ref{tlogg3293}.  
\label{tlogg3766} }
\end{figure}

\begin{figure}
\includegraphics[angle=90,scale=0.35]{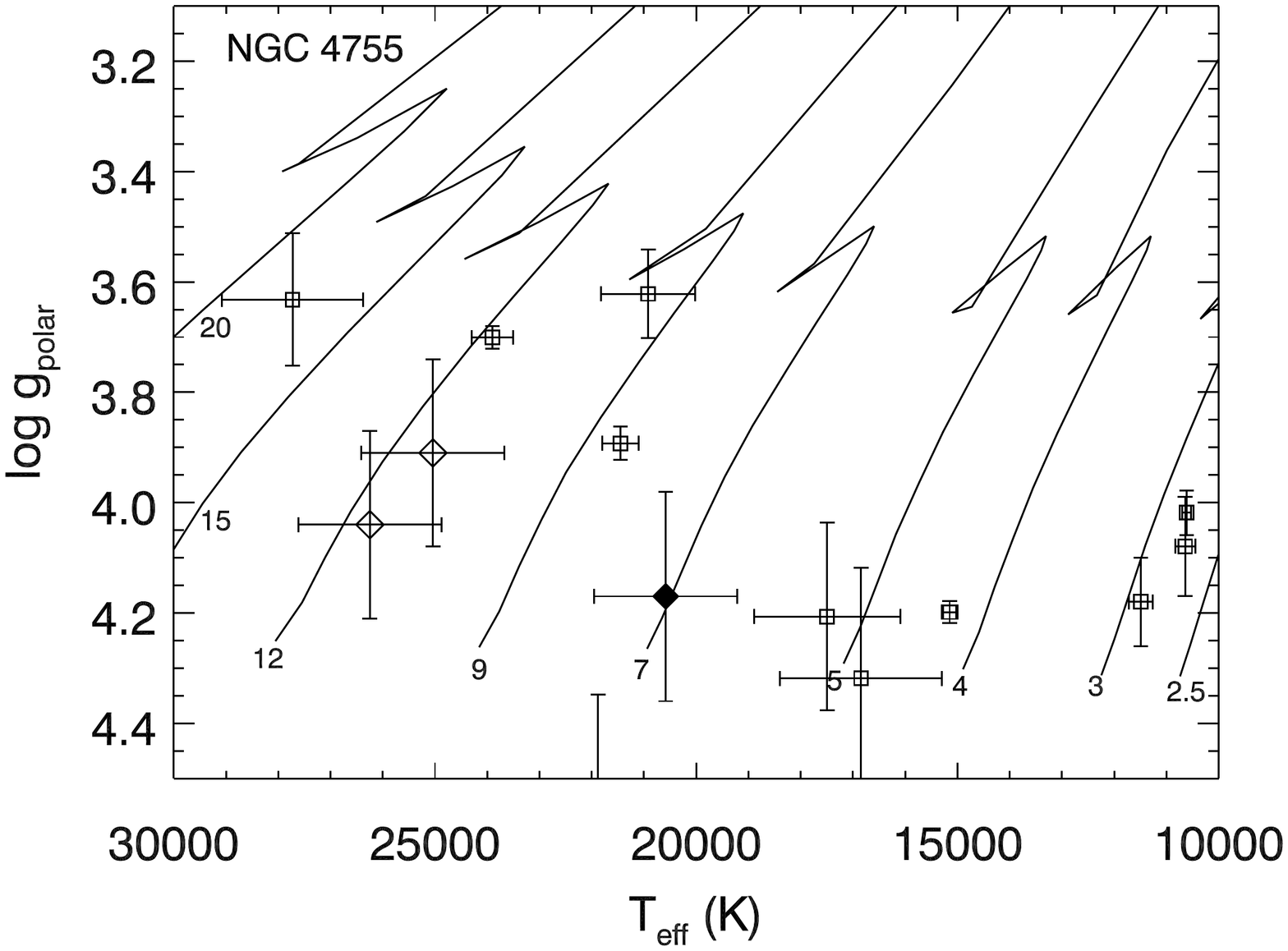}
\caption{$T_{\rm eff}$ and $\log g$ are plotted for the cluster NGC 4755 in the same format as Figure \ref{tlogg3293}.  
\label{tlogg4755} }
\end{figure}

\begin{figure}
\includegraphics[angle=90,scale=0.35]{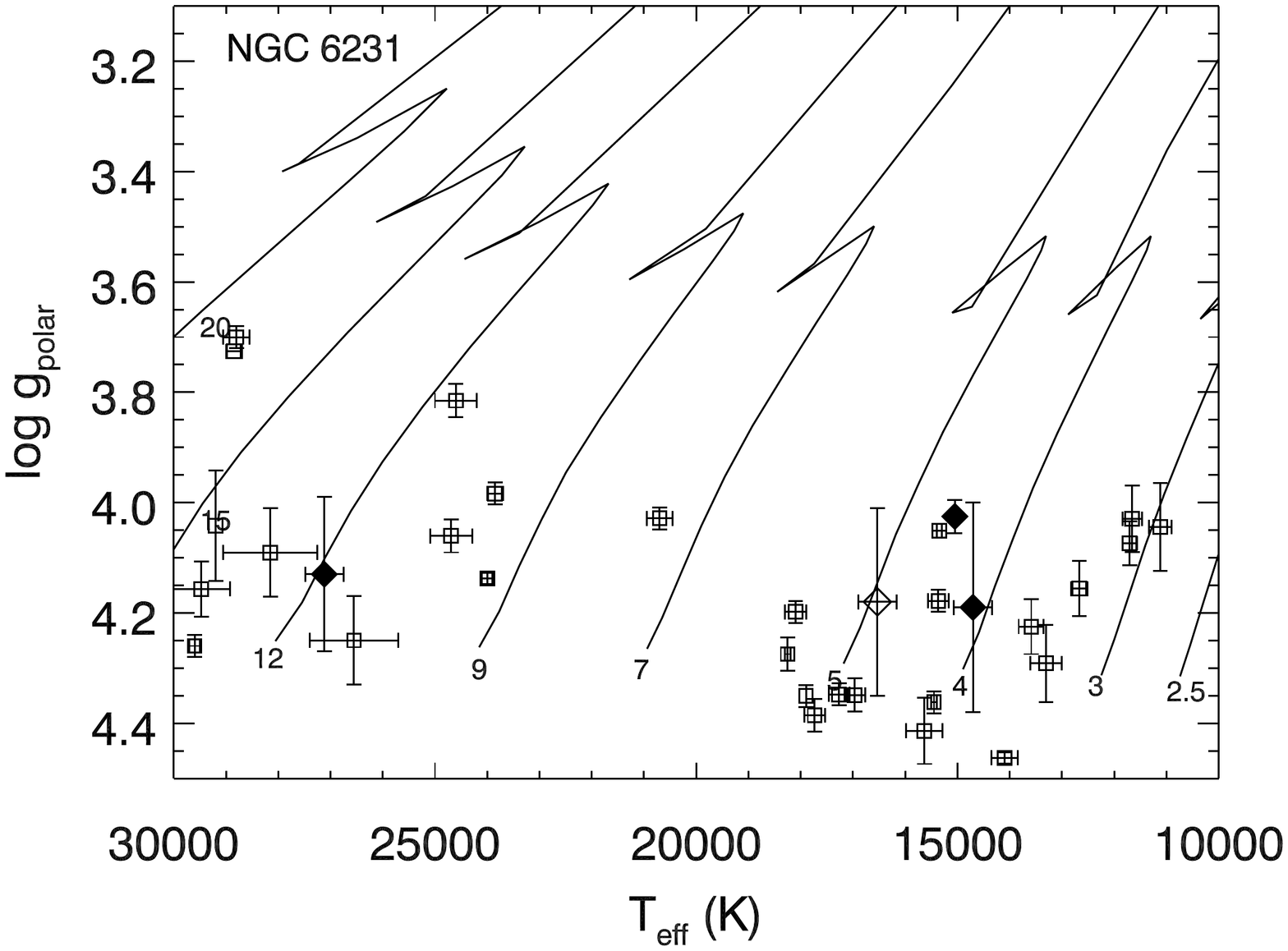}
\caption{$T_{\rm eff}$ and $\log g$ are plotted for the cluster NGC 6231 in the same format as Figure \ref{tlogg3293}.  
\label{tlogg6231} }
\end{figure}

In Figures \ref{tlogg3293}--\ref{tlogg6231}, we plot the $T_{\rm eff}$ and $\log g$ distributions for each cluster with the corresponding evolutionary tracks for nonrotating stars from \citet{schaller1992}.  

\placefigure{tlogg3293}
\placefigure{tlogg3766}
\placefigure{tlogg4755}
\placefigure{tlogg6231}

The large population of Be transients with $15000 < T_{\rm eff} < 20000$ K in NGC 3766 is striking, but we doubt that this is significant.  Due to the cluster's older age, its members are more evolved and are generally not found at higher temperatures.  In addition, nearly every Be transient in the other three clusters lies outside this range,  and the transients are well distributed among the entire observed range of temperatures.
The relationship between transients and cluster age is not clear, and it is worth investigating the transient Be star phenomenon in more clusters to determine whether there is an evolutionary dependence.  

Like the transient Be stars, the stable Be stars are equally distributed over the entire temperature range.  There is no clear dependence in $T_{\rm eff}$ and $\log g$ that distinguishes the stable Be stars from the transients or normal B-type star populations.


\section{Summary}

From H$\alpha$ spectroscopy of 296 stars in 8 open clusters, we have identified a total of 23 Be stars that experience transitions from the Be to B spectroscopic phase (including those in NGC 3766 that we reported in Paper 3).  We also find 22 Be stars that have relatively stable disks that are consistently present in our multiple epochs of H$\alpha$ spectroscopy.  Transient Be stars do not appear equally common in the open clusters, and we find higher rates of disk appearance and disappearance events in Collinder 272, Hogg 16, NGC 3766, and NGC 6231.  These clusters will be important sites to study the disk dissipation and formation mechanisms for Be stars.  

From blue optical spectroscopy and available Str\"omgren photometry of 128 Be and B-type stars in 4 open clusters, we have measured the $T_{\rm eff}$, $\log g_{\rm polar}$, and $V \sin i$ of these stars.  Our sample of physical parameters includes 17 transient Be stars and 12 stable Be stars, and we compare their rotational properties to find that all of the Be stars have a similar distribution of $V \sin i$ and are more rapidly rotating than normal B-type stars.  Other than the temporary disappearance of their disk emission, the transient Be stars do not exhibit significant physical differences from Be stars with relatively stable disks.


\acknowledgments

We are grateful to the anonymous referee for comments that greatly improved this manuscript.  We thank the staff at CTIO, especially Knut Olsen and all of the night assistants, for their hard work and patient support during our many observing runs.  Support for MVM has been provided by an NSF Astronomy and Astrophysics Postdoctoral Fellowship and by an institutional grant from Lehigh University.  This material is based upon work supported by the National Science Foundation under Grants No.\ AST-0606861, AST-0507219, and AST-0401460.

{\it Facility:} \facility{CTIO:4m (Hydra)}


\clearpage
\begin{deluxetable}{lcccccl}
\rotate
\tablewidth{0pt}
\tabletypesize{\scriptsize}
\tablecaption{Journal of Spectroscopy\label{spectroscopy}}
\tablehead{
\colhead{UT} &
\colhead{Range} &
\colhead{Resolving Power} &
\colhead{Slit Plate} &
\colhead{Grating/} &
\colhead{ } &
\colhead{ } \\
\colhead{Dates} &
\colhead{(\AA)} &
\colhead{($\lambda/\Delta\lambda$)} &
\colhead{($\mu$m)} &
\colhead{Order} &
\colhead{Filter} & 
\colhead{Targets} }
\startdata
2005 Feb $2-3$    &  $4100-6900$  &  1900  &   200   &  KPGL3/1  & \nodata &  Collinder 272, IC 2581, NGC 3293, NGC 3766, NGC 4755 \\
2006 May 13       &  $3790-4708$  &  3170  & \nodata &  KPGLD/2  &  BG39   &  NGC 3766, NGC 6231 \\
2006 May $14-15$  &  $5125-8000$  &  1560  & \nodata &  KPGL3/1  & \nodata &  Collinder 272, NGC 3293, NGC 3766, NGC 6231, NGC 6664 \\
2007 May $3-4$    &  $3790-4708$  &  4250  &   200   &  KPGLD/2  &  BG39   &  NGC 3293, NGC 3766, NGC 4755 \\
2007 May $5-6$    &  $5125-8000$  &  2000  &   200   &  KPGL3/1  & \nodata &  Collinder 272, IC 2581, NGC 3293, NGC 3766, NGC 4755, NGC 6231, NGC 6664 \\
\enddata
\end{deluxetable}

\clearpage
\begin{deluxetable}{lrrrll}
\tablewidth{0pt}
\tabletypesize{\scriptsize}
\tablecaption{H$\alpha$ Equivalent Widths\label{eqwidths}}
\tablehead{
\colhead{Cluster,} &
\colhead{HJD (t - 2,450,000),} &
\colhead{HJD (t - 2,450,000),} &
\colhead{HJD (t - 2,450,000),} &
\colhead{Emission status} &
\colhead{Emission status}  \\
\colhead{MG ID} &
\colhead{$W_{H\alpha}$ (\AA) (2005 Feb)} &
\colhead{$W_{H\alpha}$ (\AA) (2006 May)} &
\colhead{$W_{H\alpha}$ (\AA) (2007 May)} &
\colhead{in Paper 2} &
\colhead{in this work} }
\startdata
Collinder 272  & 3404.832 & 3870.178 & 4225.757  &  \nodata  &  \nodata  \\
\hline
\phn 42  &     8.60  &     8.87  &     7.35  &  \nodata  &  \nodata  \\
\phn 44  &     7.60  &     7.64  &    10.22  &  \nodata  &  \nodata  \\
 157  &     8.84  &     9.17  &    10.45  &  \nodata  &  \nodata  \\
  188  &     8.20  &     8.94  &     9.11  &  \nodata  &  \nodata  \\
  199  &    10.45  &    10.77  &    10.94  &  \nodata  &  \nodata  \\
  222  &    10.52  &    10.95  &    10.97  &  \nodata  &  \nodata  \\
  225  &     9.82  &    10.48  &     9.29  &  \nodata  &  \nodata  \\
  254  &     4.43  &     4.62  &     4.63  &  \nodata  &  \nodata  \\
  267  &     5.92  &     6.28  &     6.23  &  \nodata  &  \nodata  \\
  288  &  \nodata  &    10.54  &    10.13  &  \nodata  &  \nodata  \\
  302  &     6.54  &     6.91  &     6.77  &  \nodata  &  \nodata  \\
  309  &     6.62  &     7.31  &     7.41  &  \nodata  &  \nodata  \\
  310  &    10.66  &    10.70  &    10.85  &  \nodata  &  \nodata  \\
  311  &     7.74  &     7.74  &     7.84  &  \nodata  &  \nodata  \\
  329  &     8.37  &     9.64  &    10.00  &  \nodata  &  \nodata  \\
  349  &     7.99  &     8.65  &     8.47  &  \nodata  &  \nodata  \\
  376  &    10.79  &    10.54  &    10.67  &  \nodata  &  \nodata  \\
  377  &     7.50  &     8.40  &     8.27  &  \nodata  &  \nodata  \\
  397  &     8.93  &     9.20  &     8.58  &  \nodata  &  \nodata  \\
  406  &  \nodata  &    13.00  &    12.36  &  \nodata  &  \nodata  \\
  448  &    10.47  &     8.92  &     9.21  &  \nodata  &  \nodata  \\
  484  &    12.58  &     9.13  &  \nodata  &  \nodata  &  \nodata  \\
  508  & $-$42.93  & $-$38.08  &  \nodata  &  Be  &  Be  \\  
  513  &  $-$6.90  &  $-$2.73  &  $-$0.46  &  Be?  &  Be  \\  
  560  &    10.33  &    10.08  &    10.32  &  Be?  &  \nodata  \\  
  561  &     9.00  &     8.65  &     8.62  &  \nodata  &  \nodata  \\
  562  &  $-$4.56  &  $-$1.70  &     2.92  &  Be?  &  Be transient  \\  
  565  &    10.23  &    10.52  &  \nodata  &  Be  &  \nodata  \\  
  621  &     1.01  &     4.90  &     8.43  &  \nodata  &  Be transient  \\
  665  &     7.08  &     7.34  &     7.84  &  \nodata  &  \nodata  \\
  678  &  \nodata  &     6.79  &     7.27  &  \nodata  &  \nodata  \\
  708  &  \nodata  &    10.59  &    10.45  &  \nodata  &  \nodata  \\
  717  &     6.80  &     7.74  &     7.62  &  \nodata  &  \nodata  \\
  755  &    11.67  &    11.49  &    11.44  &  \nodata  &  \nodata  \\
  767  &     7.71  &    10.94  &     7.52  &  Be  &  Be transient  \\
  793  &  \nodata  &    11.71  &    10.97  &  \nodata  &  \nodata  \\
\\
\hline
Hogg 16        & 3404.832 & 3870.178 & 4225.757  &  \nodata  &  \nodata  \\
\hline
\phn   94  &     8.10  &     8.26  &  \nodata  &  \nodata  &  \nodata  \\
  125  &     4.44  &     4.55  &     3.93  &  \nodata  &  \nodata  \\
  134  &    12.03  &    11.61  &  \nodata  &  \nodata  &  \nodata  \\
  154  &     8.47  &     8.55  &  \nodata  &  \nodata  &  \nodata  \\
  185  &     1.69  &     5.54  &     2.69  &  \nodata  &  Be transient  \\
  209  &     4.32  &     4.54  &     4.21  &  \nodata  &  \nodata  \\
  217  &    12.16  &    12.14  &    14.34  &  \nodata  &  \nodata  \\
  227  &  \nodata  &     5.40  &  \nodata  &  \nodata  &  \nodata  \\
  254  &  \nodata  &    13.17  &  \nodata  &  \nodata  &  \nodata  \\
  317  &     4.43  &     5.05  &     4.77  &  \nodata  &  \nodata  \\
\\
\hline
IC 2581\tablenotemark{a}         & 3404.708 &  \nodata  & 4225.5494 & \nodata  \\
\hline
\phn\phn    5  &     8.72  &  \nodata  &    11.07  &  \nodata  &  \nodata  \\
\phn\phn    9  &     9.22  &  \nodata  &    10.04  &  \nodata  &  \nodata  \\
\phn   10  &     5.87  &  \nodata  &     6.05  &  Be?  &  \nodata  \\  
\phn   19  &     6.91  &  \nodata  &     7.27  &  \nodata  &  \nodata  \\
\phn   27  &    11.12  &  \nodata  &    11.77  &  \nodata  &  \nodata  \\
\phn   42  &     7.54  &  \nodata  &     7.21  &  \nodata  &  \nodata  \\
\phn   44  &     5.84  &  \nodata  &     6.40  &  \nodata  &  \nodata  \\
\phn   48  &     7.05  &  \nodata  &     7.21  &  \nodata  &  \nodata  \\
\phn   73  &    12.96  &  \nodata  &    11.93  &  \nodata  &  \nodata  \\
\phn   84  &     9.46  &  \nodata  &    10.35  &  \nodata  &  \nodata  \\
\phn   99  &    10.08  &  \nodata  &    11.34  &  \nodata  &  \nodata  \\
  105  &    10.31  &  \nodata  &    10.76  &  \nodata  &  \nodata  \\
  113  &     8.40  &  \nodata  &    10.48  &  \nodata  &  Be transient  \\
  119  &     4.05  &  \nodata  &     4.34  &  \nodata  &  \nodata  \\
  122  &     4.06  &  \nodata  &     5.65  &  \nodata  &  \nodata  \\
  123  &     5.80  &  \nodata  &     7.10  &  Be?  &  \nodata  \\  
  153  &     4.53  &  \nodata  &     4.92  &  \nodata  &  \nodata  \\
  157  &     6.52  &  \nodata  &     7.11  &  Be?  &  \nodata  \\  
  158  & $-$14.88  &  \nodata  &  \nodata  &  Be  &  Be  \\
  173  &     8.59  &  \nodata  &     8.43  &  \nodata  &  \nodata  \\
  179  &     7.24  &  \nodata  &     7.74  &  \nodata  &  \nodata  \\
  181  &     7.09  &  \nodata  &     7.30  &  \nodata  &  \nodata  \\
  188  &     6.99  &  \nodata  &     7.83  &  \nodata  &  \nodata  \\
  210  &     2.80  &  \nodata  &     3.35  &  \nodata  &  \nodata  \\
  216  &     9.10  &  \nodata  &     9.38  &  \nodata  &  \nodata  \\
  219  &     3.06  &  \nodata  &     6.90  &  Be?  &  \nodata  \\  
  233  &     7.74  &  \nodata  &     8.57  &  \nodata  &  \nodata  \\
  243  & $-$48.51  &  \nodata  & $-$46.44  &  Be  &  Be  \\
  247  &     7.07  &  \nodata  &     7.46  &  \nodata  &  \nodata  \\
  251  &     3.69  &  \nodata  &     3.97  &  \nodata  &  \nodata  \\
  267  &     0.88  &  \nodata  &     1.05  &  Be?  &  \nodata  \\  
  270  &     5.13  &  \nodata  &     6.64  &  Be?  &  \nodata  \\  
  277  &     7.54  &  \nodata  &     7.55  &  \nodata  &  \nodata  \\
  295  &     3.90  &  \nodata  &     4.07  &  \nodata  &  \nodata  \\
  298  &  \nodata  &  \nodata  &     7.21  &  \nodata  &  \nodata  \\
  314  &     8.97  &  \nodata  &     9.29  &  \nodata  &  \nodata  \\
  316  &     9.32  &  \nodata  &     9.24  &  \nodata  &  \nodata  \\
  323  &     7.25  &  \nodata  &     6.27  &  Be?  &  \nodata  \\  
  329  &     6.24  &  \nodata  &     6.34  &  \nodata  &  \nodata  \\
  370  &     9.41  &  \nodata  &     9.57  &  \nodata  &  \nodata  \\
\\
\hline
NGC 3293\tablenotemark{a} & 3404.772 & 3870.506 & 4223.321 &  \nodata  &  \nodata  \\
\hline
\phn\phn    1  &     6.02  &     5.43  &     4.92  &  \nodata  &  \nodata  \\
\phn\phn    2  &     5.47  &     5.22  &     4.68  &  \nodata  &  \nodata  \\
\phn\phn    4  &     4.24  &     3.29  &  $-$0.60  &  Be?  &  \nodata  \\  
\phn\phn    7  &     8.10  &     7.87  &     7.05  &  \nodata  &  \nodata  \\
\phn\phn    8  &     4.75  &     4.67  &     4.07  &  Be?  &  \nodata  \\  
\phn\phn    9  &     5.64  &     6.27  &  \nodata  &  \nodata  &  \nodata  \\
\phn   16  &     9.08  &     9.37  &     9.05  &  \nodata  &  \nodata  \\
\phn   18  &     9.27  &     8.85  &  \nodata  &  \nodata  &  \nodata  \\
\phn   22  &     4.35  &     4.43  &     4.50  &  \nodata  &  \nodata  \\
\phn   23  &     4.18  &     6.14  &     6.12  &  \nodata  &  \nodata  \\
\phn   27  &     6.02  &     6.26  &     6.10  &  Be?  &  \nodata  \\  
\phn   28  &     3.96  &     4.20  &     4.20  &  \nodata  &  \nodata  \\
\phn   40  &     3.51  &     4.10  &     4.16  &  \nodata  &  \nodata  \\
\phn   42  &  $-$1.60  &  $-$1.48  &  $-$2.26  &  Be  &  Be  \\
\phn   46  &     2.14  &     2.83  &     2.72  &  Be  &  Be  \\
\phn   47  &     3.32  &     3.70  &     3.86  &  \nodata  &  \nodata  \\
\phn   51  &     4.58  &     5.68  &     4.67  &  Be?  &  \nodata  \\  
\phn   54  &     6.57  &     5.06  &     4.23  &  Be?  &  \nodata  \\  
\phn   61  &     6.03  &     5.94  &     5.27  &  \nodata  &  \nodata  \\
\phn   62  &     3.52  &     3.73  &     3.88  &  \nodata  &  \nodata  \\
\phn   67  &     6.18  &     6.39  &     6.12  &  \nodata  &  \nodata  \\
\phn   74  &     2.56  &     3.83  &     4.16  &  \nodata  &  Be transient  \\
\phn   83  &     0.47  &     4.14  &     4.15  &  \nodata  &  Be transient  \\
\phn   84  &     7.22  &     7.31  &     7.34  &  Be?  &  \nodata  \\  
\phn   90  &     4.09  &     4.08  &     4.27  &  \nodata  &  \nodata  \\
\phn   95  &     3.20  &     3.84  &     3.97  &  \nodata  &  \nodata  \\
\phn   99  & $-$11.99  & $-$15.89  & $-$20.47  &  Be?  &  Be  \\
  100  &     3.35  &     3.26  &     3.65  &  \nodata  &  \nodata  \\
  107  &     7.74  &     8.78  &     9.01  &  \nodata  &  \nodata  \\
  110  &     3.89  &     5.19  &     5.23  &  \nodata  &  \nodata  \\
  120  &     5.57  &     5.66  &     5.63  &  \nodata  &  \nodata  \\
\\
\hline
NGC 3766  &   3403.790 &  3870.224  & 4225.662 &  \nodata  &  \nodata  \\
\hline
\phn\phn    2  &     3.24  &     4.07  &     4.11  &  \nodata  &  \nodata  \\
\phn\phn    8  &  \nodata  &  \nodata  &     6.44  &  \nodata  &  \nodata  \\
\phn   16  &     4.28  &     4.26  &     4.39  &  \nodata  &  \nodata  \\
\phn   23  &     9.22  &     9.37  &     9.17  &  \nodata  &  \nodata  \\
\phn   25  &     3.18  &     5.30  &     1.54  &  \nodata  &  Be transient \\
\phn   26  &  \nodata  &  \nodata  &     5.46  &  \nodata  &  \nodata  \\
\phn   27  &     5.38  &     5.69  &     5.59  &  \nodata  &  \nodata  \\
\phn   28  &  \nodata  &  \nodata  &     6.38  &  \nodata  &  \nodata  \\
\phn   29  &  \nodata  &  \nodata  &     4.47  &  \nodata  &  \nodata  \\
\phn   31  &     3.99  &     0.57  &  $-$6.42  &  \nodata  &  Be transient \\
\phn   33  &  \nodata  &  \nodata  &     4.86  &  \nodata  &  \nodata  \\
\phn   36  &     8.46  &     8.56  &     8.43  &  \nodata  &  \nodata  \\
\phn   40  &  \nodata  &  \nodata  &     5.80  &  \nodata  &  \nodata  \\
\phn   41  &     4.88  &     5.28  &     5.02  &  \nodata  &  \nodata  \\
\phn   42  &     6.07  &     6.02  &     5.85  &  \nodata  &  \nodata  \\
\phn   45  &     4.98  &     5.10  &     4.35  &  \nodata  &  \nodata  \\
\phn   46  &  \nodata  &  \nodata  &     5.49  &  \nodata  &  \nodata  \\
\phn   47  & $-$11.33  & $-$11.29  &  $-$6.43  &  Be  &  Be  \\
\phn   49  &     4.37  &     4.89  &     4.70  &  \nodata  &  \nodata  \\
\phn   52  &  \nodata  &  \nodata  &     4.61  &  \nodata  &  \nodata  \\
\phn   54  &     5.18  &     5.61  &     5.31  &  \nodata  &  \nodata  \\
\phn   55  &     5.03  &     5.12  &     5.25  &  \nodata  &  \nodata  \\
\phn   57  &     5.67  &     5.86  &     5.66  &  \nodata  &  \nodata  \\
\phn   61  &     4.09  &     4.20  &     4.24  &  Be?  &  \nodata  \\
\phn   69  &     2.89  &     3.19  &     3.16  &  \nodata  &  \nodata  \\
\phn   72  &     8.27  &     8.65  &     7.08  &  \nodata  &  \nodata  \\
\phn   73  &     0.42  &     3.28  &     0.06  &  Be?  &  Be transient  \\
\phn   77  &     4.61  &     5.72  &     5.50  &  \nodata  &  \nodata  \\
\phn   83  &     4.07  &     0.84  &     2.93  &  Be?  &  Be transient  \\
\phn   87  &  \nodata  &  \nodata  &     5.35  &  \nodata  &  \nodata  \\
\phn   89  &  \nodata  &  \nodata  &     5.70  &  \nodata  &  \nodata  \\
\phn   92  &     2.31  &     2.73  &     3.79  &  Be?  &  Be transient  \\
\phn   94  &     5.94  &     5.91  &     5.86  &  Be?  &  \nodata  \\
\phn   96  &     6.39  &     6.87  &     7.53  &  \nodata  &  \nodata  \\
\phn   98  &     3.10  &     0.21  &     4.14  &  \nodata  &  Be transient  \\
 101  &     7.40  &     7.44  &  \nodata  &  \nodata  &  \nodata  \\
 103  &  \nodata  &  \nodata  &     5.21  &  \nodata  &  \nodata  \\
 109  &  \nodata  &  \nodata  &     5.10  &  \nodata  &  \nodata  \\
 110  &  \nodata  &  \nodata  &     4.83  &  \nodata  &  \nodata  \\
 111  &  \nodata  &  \nodata  &     4.56  &  \nodata  &  \nodata  \\
 118  &     7.40  &     7.03  &     6.73  &  \nodata  &  \nodata  \\
 119  &  $-$5.48  &  $-$3.19  &  $-$8.67  &  \nodata  &  Be transient  \\
 121  &  \nodata  &  \nodata  &     6.54  &  \nodata  &  \nodata  \\
 126  &     7.46  &     8.49  &     7.74  &  Be?  &  \nodata  \\
 127  &  $-$4.95  &  $-$4.00  &  $-$5.03  &  Be  &  Be  \\
 129  &     8.66  &     7.94  &     7.96  &  \nodata  &  \nodata  \\
 130  &     2.00  &     5.04  &     5.09  &  Be?  &  Be transient  \\
 131  &  \nodata  &  \nodata  &     5.30  &  \nodata  &  \nodata  \\
 133  &  $-$1.15  &  $-$9.14  & $-$12.72  &  \nodata  &  Be transient  \\
 137  &  \nodata  &  \nodata  &     6.02  &  \nodata  &  \nodata  \\
 138  &  \nodata  &  \nodata  &     5.09  &  \nodata  &  \nodata  \\
 139  &     5.46  &     4.24  &     5.05  &  Be?  &  Be transient  \\
 142  &  \nodata  &  \nodata  &     6.57  &  \nodata  &  \nodata  \\
 145  &  \nodata  &  \nodata  &     7.43  &  \nodata  &  \nodata  \\
 148  &  \nodata  &  \nodata  &     4.84  &  \nodata  &  \nodata  \\
 149  &  \nodata  &  \nodata  &     7.55  &  \nodata  &  \nodata  \\
 150  &  \nodata  &  \nodata  &     6.83  &  \nodata  &  \nodata  \\
 154  & $-$40.07  & $-$33.56  & $-$33.40  &  Be  &  Be  \\
 155  &     7.92  &     7.90  &     7.42  &  \nodata  &  \nodata  \\
 161  &     3.92  &     4.19  &     4.18  &  \nodata  &  \nodata  \\
 162  &     9.15  &     9.23  &     8.85  &  \nodata  &  \nodata  \\
  164  &  \nodata  &  \nodata  &     6.46  &  \nodata  &  \nodata  \\
  170  &     4.81  &     5.26  &     5.34  &  \nodata  &  \nodata  \\
  171  &  \nodata  &  \nodata  &     5.11  &  \nodata  &  \nodata  \\
  173  &     7.79  &     7.92  &     7.80  &  \nodata  &  \nodata  \\
  175  &     6.67  &     6.74  &     6.64  &  \nodata  &  \nodata  \\
  176  &  \nodata  &  \nodata  &     3.54  &  \nodata  &  \nodata  \\
  178  &     6.32  &     7.71  &     7.40  &  \nodata  &  \nodata  \\
  179  &  \nodata  &  \nodata  &     4.25  &  \nodata  &  \nodata  \\
  184  &  \nodata  &  \nodata  &     5.72  &  \nodata  &  \nodata  \\
  190  &     6.81  &     6.86  &  \nodata  &  \nodata  &  \nodata  \\
  192  &  \nodata  &  \nodata  &     6.87  &  \nodata  &  \nodata  \\
  196  &     4.32  &     4.39  &     1.69  &  \nodata  &  Be transient  \\
  197  &     7.65  &     8.85  &     8.49  &  \nodata  &  \nodata  \\
  198  & $-$43.40  & $-$53.71  & $-$53.96  &  Be  &  Be  \\
  200  &  $-$6.66  &  $-$4.62  &  $-$5.08  &  Be  &  Be  \\
\\
\hline
NGC 4755       & 3403.863 &  \nodata  & 4223.717 &  \nodata  &  \nodata  \\
\hline
\phn\phn    1  &     6.31  &  \nodata  &     6.63  &  \nodata  &  \nodata  \\
\phn\phn    6  &     9.50  &  \nodata  &    10.33  &  \nodata  &  \nodata  \\
\phn   15  &     4.53  &  \nodata  &     4.96  &  \nodata  &  \nodata  \\
\phn   18  &     7.42  &  \nodata  &     2.57  &  \nodata  &  Be transient  \\
\phn   25  &     3.59  &  \nodata  &     4.08  &  \nodata  &  \nodata  \\
\phn   27  &     9.94  &  \nodata  &    11.11  &  \nodata  &  \nodata  \\
\phn   30  &     9.61  &  \nodata  &  \nodata  &  \nodata  &  \nodata  \\
\phn   31  &     4.85  &  \nodata  &  $-$2.09  &  \nodata  &  Be transient  \\
\phn   43  &     7.86  &  \nodata  &  \nodata  &  Be  &  \nodata  \\  
\phn   45  &     6.02  &  \nodata  &     6.51  &  \nodata  &  \nodata  \\
\phn   47  &     5.48  &  \nodata  &     6.42  &  \nodata  &  \nodata  \\
\phn   53  &     2.31  &  \nodata  &     2.69  &  Be?  &  \nodata  \\  
\phn   54  &     9.35  &  \nodata  &    10.06  &  \nodata  &  \nodata  \\
\phn   71  &     5.01  &  \nodata  &     5.80  &  \nodata  &  \nodata  \\
\phn   73  &     8.33  &  \nodata  &     9.11  &  \nodata  &  \nodata  \\
\phn   75  &     5.95  &  \nodata  &     6.56  &  \nodata  &  \nodata  \\
\phn   93  &     5.85  &  \nodata  &     6.79  &  \nodata  &  \nodata  \\
  103  &     4.26  &  \nodata  &     5.35  &  \nodata  &  \nodata  \\
  120  &     5.20  &  \nodata  &     5.99  &  \nodata  &  \nodata  \\
  121  &     4.42  &  \nodata  &     6.23  &  \nodata  &  \nodata  \\
  129  & $-$48.80  &  \nodata  & $-$53.35  &  Be  &  Be  \\
  131  &     6.03  &  \nodata  &  \nodata  &  Be?  &  \nodata  \\ 
  144  &     8.54  &  \nodata  &     9.00  &  \nodata  &  \nodata  \\
  148  &     6.48  &  \nodata  &     6.99  &  \nodata  &  \nodata  \\
  152  &    10.70  &  \nodata  &    11.37  &  \nodata  &  \nodata  \\
  166  &    10.81  &  \nodata  &    11.60  &  Be?  &  \nodata  \\  
  168  &     9.51  &  \nodata  &  \nodata  &  Be?  &  \nodata  \\
  169  &     4.99  &  \nodata  &  \nodata  &  \nodata  &  \nodata  \\
  171  & $-$55.57  &  \nodata  & $-$55.68  &  Be  &  Be  \\
  172  &     3.50  &  \nodata  &     4.20  &  \nodata  &  \nodata  \\
  180  & $-$44.23  &  \nodata  & $-$36.48  &  Be  &  Be  \\
  183  &     4.11  &  \nodata  &     4.52  &  \nodata  &  \nodata  \\
  187  & $-$14.19  &  \nodata  & $-$13.86  &  Be  &  Be  \\
  195  &     2.47  &  \nodata  &     3.12  &  \nodata  &  \nodata  \\
  205  &     5.53  &  \nodata  &  \nodata  &  \nodata  &  \nodata  \\
  215  &     5.58  &  \nodata  &  \nodata  &  \nodata  &  \nodata  \\
  222  &     8.27  &  \nodata  &     8.23  &  Be?  &  \nodata  \\ 
  223  &    11.81  &  \nodata  &    11.08  &  Be?  &  \nodata  \\  
  252  &     5.08  &  \nodata  &  \nodata  &  \nodata  &  \nodata  \\
  261  &     8.13  &  \nodata  &     8.47  &  \nodata  &  \nodata  \\
\\
\hline
NGC 6231        &  \nodata  & 3869.828 & 4223.813 &  \nodata  &  \nodata  \\
\hline
\phn\phn    2  &  \nodata  &     5.29  &     4.78  &  Be?  &  Be transient  \\ 
\phn\phn    4  &  \nodata  &     8.42  &     8.14  &  \nodata  &  \nodata  \\
\phn   15  &  \nodata  &     6.44  &     5.94  &  \nodata  &  \nodata  \\
\phn   32  &  \nodata  &     4.37  &     4.38  &  \nodata  &  \nodata  \\
\phn   38  &  \nodata  &     8.38  &     8.20  &  \nodata  &  \nodata  \\
\phn   45  &  \nodata  &     5.75  &     5.53  &  \nodata  &  \nodata  \\
\phn   55  &  \nodata  &     5.62  &     5.35  &  \nodata  &  \nodata  \\
\phn   57  &  \nodata  &     8.42  &     7.77  &  \nodata  &  \nodata  \\
\phn   63  &  \nodata  &     7.15  &     7.16  &  \nodata  &  \nodata  \\
\phn   67  &  \nodata  &     6.05  &     5.59  &  Be  &  \nodata  \\
\phn   79  &  \nodata  &  $-$5.46  &  \nodata  &  Be  &  Be  \\
\phn   86  &  \nodata  &     8.24  &     7.43  &  \nodata  &  \nodata  \\
\phn   90  &  \nodata  &     3.82  &     3.17  &  \nodata  &  Be transient  \\
\phn   91  &  \nodata  &     5.18  &     4.97  &  \nodata  &  \nodata  \\
\phn   92  &  \nodata  &     9.82  &     8.32  &  \nodata  &  \nodata  \\
  110  &  \nodata  &     4.57  &     4.48  &  \nodata  &  \nodata  \\
  111  &  \nodata  &     4.27  &     4.22  &  \nodata  &  \nodata  \\
  114  &  \nodata  &     6.48  &     5.74  &  \nodata  &  \nodata  \\
  116  &  \nodata  &     4.46  &     4.49  &  \nodata  &  \nodata  \\
  121  &  \nodata  &     7.52  &     7.04  &  \nodata  &  \nodata  \\
  124  &  \nodata  &     4.49  &     4.43  &  \nodata  &  \nodata  \\
  132  &  \nodata  &     3.98  &     4.03  &  \nodata  &  \nodata  \\
  137  &  \nodata  &     4.12  &     4.04  &  \nodata  &  \nodata  \\
  156  &  \nodata  &     8.94  &     8.38  &  \nodata  &  \nodata  \\
  162  &  \nodata  &     3.14  &     3.23  &  \nodata  &  \nodata  \\
  165  &  \nodata  &     3.27  &     3.39  &  \nodata  &  \nodata  \\
  168  &  \nodata  &     3.94  &     3.95  &  \nodata  &  \nodata  \\
  169  &  \nodata  &     6.34  &     5.99  &  \nodata  &  \nodata  \\
  178  &  \nodata  &     6.87  &     6.98  &  \nodata  &  \nodata  \\
  185  &  \nodata  &     7.82  &     7.12  &  \nodata  &  \nodata  \\
  190  &  \nodata  &     3.52  &     3.49  &  \nodata  &  \nodata  \\
  213  &  \nodata  &     6.32  &     6.19  &  \nodata  &  \nodata  \\
  218  &  \nodata  &     4.14  &     4.11  &  \nodata  &  \nodata  \\
  221  &  \nodata  &     6.71  &     6.96  &  Be?  &  Be transient  \\
  235  &  \nodata  &     7.06  &     6.97  &  \nodata  &  \nodata  \\
  245  &  \nodata  &     5.65  &     5.67  &  \nodata  &  \nodata  \\
  268  &  \nodata  &     7.49  &     7.96  &  \nodata  &  \nodata  \\
\\
\hline
NGC 6664       &  \nodata  & 3870.933 & 4223.930 &  \nodata  &  \nodata  \\
\hline
\phn\phn    5  &  \nodata  &  \nodata  &    11.20  &  \nodata  &  \nodata  \\
\phn   13  &  \nodata  &     6.88  &     6.82  &  \nodata  &  \nodata  \\
\phn   33  &  \nodata  &     8.09  &     7.84  &  \nodata  &  \nodata  \\
\phn   35  &  \nodata  &     6.55  &     6.28  &  \nodata  &  \nodata  \\
\phn   45  &  \nodata  &    10.20  &    10.32  &  \nodata  &  \nodata  \\
\phn   53  &  \nodata  &     8.19  &     8.26  &  \nodata  &  \nodata  \\
\phn   62  &  \nodata  &     7.24  &     7.12  &  \nodata  &  \nodata  \\
\phn   68  &  \nodata  &     8.90  &     8.89  &  \nodata  &  \nodata  \\
\phn   73  &  \nodata  &     5.82  &     5.79  &  \nodata  &  \nodata  \\
\phn   79  &  \nodata  &     3.82  &     3.71  &  \nodata  &  \nodata  \\
\phn   81  &  \nodata  &    12.57  &    12.17  &  \nodata  &  \nodata  \\
  119  &  \nodata  &     6.42  &     6.14  &  \nodata  &  \nodata  \\
  126  &  \nodata  &     1.76  &     2.84  &  \nodata  &  Be  \\
  137  &  \nodata  &  $-$4.32  &  $-$4.37  &  Be?  &  Be  \\
  141  &  \nodata  &    11.41  &    11.15  &  \nodata  &  \nodata  \\
  143  &  \nodata  &     9.50  &     9.00  &  \nodata  &  \nodata  \\
  158  &  \nodata  &    11.35  &    11.45  &  \nodata  &  \nodata  \\
  168  &  \nodata  &     7.82  &     7.59  &  \nodata  &  \nodata  \\
  177  &  \nodata  &  $-$7.27  &  $-$7.37  &  Be  &  Be  \\
  182  &  \nodata  &     1.47  &     1.74  &  Be  &  Be  \\
  183  &  \nodata  &     7.20  &     7.21  &  \nodata  &  \nodata  \\
  207  &  \nodata  &     6.85  &     6.73  &  \nodata  &  \nodata  \\
  219  &  \nodata  &     5.69  &     5.56  &  \nodata  &  \nodata  \\
  221  &  \nodata  & $-$18.49  & $-$18.68  &  Be  &  Be  \\
  229  &  \nodata  &  \nodata  &    10.48  &  \nodata  &  \nodata  \\
  237  &  \nodata  &     7.13  &     6.84  &  \nodata  &  \nodata 
\enddata
\tablenotetext{a}{$W_{H\alpha}$ may be contaminated by nebular emission in NGC 3293 and IC 2581.}
\end{deluxetable}

\begin{deluxetable}{lccc}
\tablewidth{0pt}
\tablecaption{Fraction of Be Transients\label{transient}}
\tablehead{
\colhead{ } &
\colhead{Number of Be } &
\colhead{Number of Transient } &
\colhead{Ratio of Transients} \\
\colhead{Cluster} &
\colhead{Stars, $n$} &
\colhead{Be Stars, $t$} &
\colhead{to Total Be, $p$ (\%)} }
\startdata
Collinder 272/Hogg 16   & \phn 6  & \phn 4  &   $67 \pm 19$  \\
IC 2581                 & \phn 3  & \phn 1  &   $33 \pm 27$  \\
NGC 3293                & \phn 5  & \phn 2  &   $40 \pm 22$  \\
NGC 3766                &     16  &     11    &  $69 \pm 12$  \\
NGC 4755                & \phn 6  & \phn 2  &   $33 \pm 19$  \\
NGC 6231                & \phn 4  & \phn 3  &   $75 \pm 22$   \\
NGC 6664                & \phn 5  & \phn 0  & \phn 0 $\pm$ 18  \\
\enddata
\end{deluxetable}

\begin{deluxetable}{lccccl}
\tablewidth{0pt}
\tabletypesize{\scriptsize}
\tablecaption{Cluster Parameters \label{reddening}}
\tablehead{
\colhead{ } &
\colhead{Photometric fitting} &
\colhead{$E(b-y)$ } &
\colhead{$(V-M_V)_0$ } &
\colhead{$\log$ age} &
\colhead{ } \\
\colhead{Cluster} &
\colhead{constant, $C$ (mag)} &
\colhead{(mag)} &
\colhead{(mag)} &
\colhead{(yrs)} & 
\colhead{Reference} }
\startdata
Collinder 272           & $0.49 \pm 0.06$  &  0.34  & 11.55   &  7.11  &  Paper 2  \\
Hogg 16                 & $0.43 \pm 0.08$  &  0.31  & 11.00   &  7.05  &  Paper 2  \\
IC 2581                 & $0.51 \pm 0.05$  &  0.31  & 12.00   &  7.14  &  Paper 2  \\
NGC 3293                & $0.31 \pm 0.13$  &  0.19  &  12.24  &  7.0   &  \citet{dufton2006}  \\
NGC 3766                & $0.27 \pm 0.02$  &  0.17  &  11.6   &  7.16  &  Paper 2, Paper 3  \\
NGC 4755                & $0.43 \pm 0.02$  &  0.28  &  11.61  &  7.0   &  \citet{dufton2006}  \\
NGC 6231                & $0.44 \pm 0.06$  &  0.35  &  11.07  &  6.9   &  \citet{perry1991}, \citet{sana2006}  \\
NGC 6664                & $0.67 \pm 0.06$  &  0.56  &  10.70  &  7.66  &  Paper 2
\enddata
\end{deluxetable}

\clearpage
\begin{deluxetable}{lcccccccccccccl}
\tablewidth{0pt}
\rotate
\tabletypesize{\scriptsize}
\tablecaption{Physical Parameters of Cluster Members\label{params} }
\tablehead{
\colhead{ } &
\colhead{MG} &
\colhead{$V \sin i$} &
\colhead{$\Delta V \sin i$} &
\colhead{$T_{\rm eff}$} &
\colhead{$\Delta T_{\rm eff}$} &
\colhead{ } &
\colhead{ } &
\colhead{ } &
\colhead{$M_\star$} &
\colhead{$\Delta M_\star$} &
\colhead{$R_\star$} &
\colhead{$\Delta R_\star$} &
\colhead{$V_{\rm crit}$} &
\colhead{ }  \\
\colhead{Cluster} &
\colhead{ID} &
\colhead{(km~s$^{-1}$)} &
\colhead{(km~s$^{-1}$)} &
\colhead{(K)} &
\colhead{(K)} &
\colhead{$\log g$} &
\colhead{$\Delta \log g$} &
\colhead{$\log g_{\rm polar}$} &
\colhead{($M_\odot$)} &
\colhead{($M_\odot$)} &
\colhead{($R_\odot$)} &
\colhead{($R_\odot$)} &
\colhead{(km~s$^{-1}$)} &
\colhead{Comment}  }
\startdata
B stars: \\
\hline
NGC3293
  &    1  &  139  &  21  &  22100  &   300  &  4.72  &  0.03  &  4.74  &   6.2  &   0.2  &   1.8  &   1.6  &  668  &  \nodata  \\
  &    2  &  229  &  50  &  16528  &   400  &  4.29  &  0.05  &  4.41  &   4.5  &   0.2  &   2.2  &   0.5  &  512  &  \nodata  \\
  &    4  &  298  &  35  &  15050  &    50  &  4.15  &  0.03  &  4.37  &   3.9  &   0.0  &   2.2  &   0.3  &  482  &  \nodata  \\
  &    8  &  309  &  75  &  16636  &   650  &  3.67  &  0.08  &  4.00  &   5.5  &   0.4  &   3.9  &   0.8  &  424  &  \nodata  \\
  &    9  &  266  &  31  &  16218  &   550  &  4.17  &  0.08  &  4.35  &   4.5  &   0.2  &   2.4  &   0.4  &  493  &  \nodata  \\
  &   16  &  180  &  24  &  10665  &   132  &  4.11  &  0.07  &  4.21  &   2.6  &   0.0  &   2.1  &   0.2  &  397  &  \nodata  \\
  &   22  &  126  &   5  &  22750  &   150  &  3.75  &  0.02  &  3.84  &  10.0  &   0.2  &   6.3  &   1.8  &  448  &  \nodata  \\
  &   27  &  185  &  28  &  16640  &   250  &  4.17  &  0.02  &  4.28  &   4.8  &   0.1  &   2.6  &   0.1  &  482  &  SB2? (this work) \\
  &   28  &   54  &   6  &  22944  &   350  &  3.73  &  0.02  &  3.75  &  10.8  &   0.3  &   7.2  &   1.8  &  434  &  \nodata  \\  
  &   40  &  154  &   7  &  23375  &   300  &  3.84  &  0.02  &  3.94  &  10.0  &   0.3  &   5.6  &   2.2  &  475  &  \nodata  \\
  &   47  &  105  &   9  &  29833  &   200  &  3.91  &  0.02  &  3.91  &  16.8  &   0.3  &   7.5  &   4.6  &  532  &  \nodata  \\
  &   51  &  231  &  45  &  16280  &  1100  &  4.50  &  0.10  &  4.59  &   4.1  &   0.4  &   1.7  &   1.2  &  553  &  \nodata  \\
  &   54  &  315  &  28  &  15166  &   150  &  4.18  &  0.03  &  4.40  &   3.9  &   0.1  &   2.1  &   0.4  &  490  &  \nodata  \\
  &   61  &  297  &  54  &  15750  &   450  &  4.29  &  0.05  &  4.47  &   4.1  &   0.2  &   1.9  &   0.7  &  516  &  \nodata  \\
  &   67  &  214  &  14  &  15733  &   200  &  4.13  &  0.02  &  4.27  &   4.4  &   0.1  &   2.6  &   0.9  &  469  &  \nodata  \\
  &   84  &   70  &  10  &  19600  &   300  &  4.75  &  0.00  &  4.76  &   5.2  &   0.2  &   1.6  &   1.4  &  649  &  \nodata  \\
  &   90  &   62  &  13  &  23450  &   450  &  3.68  &  0.02  &  3.71  &  11.4  &   0.3  &   7.8  &   1.8  &  430  &  \nodata  \\
  &   95  &   77  &  10  &  27822  &   900  &  3.93  &  0.08  &  3.95  &  13.9  &   0.8  &   6.5  &   3.6  &  519  & \nodata  \\
  &  100  &  106  &  10  &  27173  &   950  &  3.78  &  0.07  &  3.83  &  14.2  &   1.0  &   7.6  &   3.3  &  486  &  \nodata  \\
  &  107  &  271  &  44  &  10967  &   171  &  3.82  &  0.07  &  4.08  &   2.9  &   0.1  &   2.6  &   0.1  &  377  &  \nodata  \\
  &  110  &  163  &   8  &  19366  &   200  &  4.02  &  0.03  &  4.13  &   6.6  &   0.1  &   3.7  &   1.4  &  477  &  \nodata  \\
  &  120  &  313  &  30  &  15650  &   250  &  3.97  &  0.02  &  4.24  &   4.5  &   0.1  &   2.7  &   0.9  &  461  &  \nodata  \\
\\
NGC3766
  &    2  &  307  &   4  &  17071  &   450  &  3.10  &  0.05  &  3.57  &   7.0  &   0.3  &   7.2  &   0.6  &  352  &  \nodata  \\
  &   16  &  144  &   7  &  19420  &   300  &  3.48  &  0.02  &  3.62  &   8.6  &   0.2  &   7.5  &   0.5  &  380  &  \nodata  \\
  &   27  &  287  &   6  &  16033  &   200  &  3.83  &  0.02  &  4.10  &   4.9  &   0.1  &   3.3  &   0.8  &  436  &  \nodata  \\
  &   29  &  288  &  12  &  20140  &   950  &  4.66  &  0.10  &  4.72  &   5.6  &   0.4  &   1.7  &   1.7  &  645  &  \nodata  \\
  &   33  &  257  &  27  &  16828  &   350  &  4.10  &  0.05  &  4.28  &   4.9  &   0.1  &   2.6  &   0.2  &  484  &  \nodata  \\
  &   36  &  283  &  22  &  11931  &   145  &  4.02  &  0.05  &  4.27  &   3.0  &   0.0  &   2.1  &   0.1  &  424  &  \nodata  \\
  &   41  &   84  &   7  &  17900  &   300  &  3.78  &  0.03  &  3.84  &   6.6  &   0.1  &   5.1  &   0.8  &  406  &  \nodata  \\
  &   42  &  239  &   5  &  16100  &   100  &  3.90  &  0.02  &  4.11  &   4.9  &   0.0  &   3.2  &   0.8  &  438  &  \nodata  \\
  &   45  &   90  &  28  &  15500  &   550  &  4.61  &  0.02  &  4.64  &   3.6  &   0.3  &   1.5  &   1.0  &  554  &  \nodata  \\
  &   46  &  272  &  27  &  15530  &   550  &  4.52  &  0.03  &  4.64  &   3.7  &   0.3  &   1.5  &   1.0  &  554  &  \nodata  \\
  &   49  &  252  &   9  &  17400  &   100  &  3.75  &  0.02  &  3.99  &   6.0  &   0.1  &   4.1  &   0.9  &  431  &  \nodata  \\
  &   54  &  192  &   9  &  17340  &   250  &  3.82  &  0.02  &  3.99  &   5.9  &   0.1  &   4.1  &   0.9  &  430  &  \nodata  \\
  &   55  &  121  &   6  &  18000  &   200  &  3.78  &  0.03  &  3.88  &   6.6  &   0.1  &   4.9  &   0.8  &  413  &  \nodata  \\
  &   57  &  238  &   8  &  16760  &   250  &  3.98  &  0.03  &  4.17  &   5.1  &   0.2  &   3.1  &   1.0  &  459  &  \nodata  \\
  &   61  &  331  &   8  &  18883  &   350  &  3.23  &  0.02  &  3.71  &   7.7  &   0.3  &   6.4  &   0.6  &  390  &  SB2? (Paper 3) \\
  &   77  &  338  &  10  &  16000  &   200  &  3.80  &  0.05  &  4.15  &   4.8  &   0.1  &   3.1  &   0.9  &  445  &  \nodata  \\
  &   89  &  189  &  36  &  11714  &   665  &  4.33  &  0.21  &  4.39  &   2.7  &   0.2  &   1.7  &   0.9  &  447  &  \nodata  \\
  &   94  &  177  &   4  &  15650  &   150  &  3.78  &  0.03  &  3.94  &   5.0  &   0.1  &   4.0  &   0.5  &  401  &  \nodata  \\
  &   96  &  121  &   5  &  13937  &   171  &  4.31  &  0.04  &  4.37  &   3.6  &   0.1  &   2.0  &   0.3  &  472  &  \nodata  \\
  &  101  &  309  &  14  &  12753  &   192  &  3.84  &  0.05  &  4.14  &   3.5  &   0.1  &   2.6  &   0.4  &  409  &  \nodata  \\
  &  109  &   86  &   6  &  17100  &   200  &  4.22  &  0.03  &  4.26  &   5.0  &   0.1  &   2.7  &   1.1  &  483  &  \nodata  \\
  &  110  &   50  &  13  &  16060  &   750  &  4.50  &  0.08  &  4.51  &   4.2  &   0.3  &   1.9  &   0.9  &  531  &  \nodata  \\
  &  111  &   66  &   9  &  18436  &   450  &  4.25  &  0.05  &  4.28  &   5.7  &   0.2  &   2.9  &   0.2  &  503  &  \nodata  \\
  &  118  &  242  &   9  &  13353  &   191  &  3.89  &  0.05  &  4.11  &   3.7  &   0.1  &   2.8  &   0.5  &  409  &  \nodata  \\
  &  126  &   83  &   9  &  12399  &   318  &  3.79  &  0.09  &  3.87  &   3.8  &   0.1  &   3.7  &   0.4  &  358  &  \nodata  \\
  &  129  &   97  &  27  &  12034  &   203  &  3.77  &  0.06  &  3.87  &   3.7  &   0.1  &   3.7  &   0.3  &  355  &  \nodata  \\
  &  131  &  250  &  23  &  12817  &   947  &  4.58  &  0.27  &  4.69  &   2.6  &   0.3  &   1.2  &   1.9  &  526  &  \nodata  \\
  &  155  &  231  &  11  &  12694  &   153  &  4.05  &  0.04  &  4.22  &   3.3  &   0.1  &   2.3  &   0.5  &  425  &  \nodata  \\
  &  161  &   73  &   7  &  18400  &   250  &  3.43  &  0.02  &  3.49  &   8.3  &   0.2  &   8.5  &   0.4  &  351  &  \nodata  \\
  &  162  &  194  &  11  &  11513  &   148  &  3.99  &  0.05  &  4.12  &   3.0  &   0.1  &   2.5  &   0.2  &  390  &  \nodata  \\
  &  170  &   65  &   6  &  18060  &   250  &  3.78  &  0.03  &  3.82  &   6.8  &   0.1  &   5.3  &   0.7  &  403  &  \nodata  \\
  &  173  &  171  &   7  &  14210  &   192  &  4.21  &  0.04  &  4.30  &   3.8  &   0.1  &   2.3  &   0.1  &  459  &  \nodata  \\
  &  175  &  252  &  13  &  13823  &   193  &  4.01  &  0.05  &  4.21  &   3.8  &   0.1  &   2.5  &   0.6  &  435  &  \nodata  \\
  &  176  &   54  &   7  &  21435  &   550  &  4.16  &  0.05  &  4.18  &   7.6  &   0.3  &   3.7  &   1.9  &  510  &  \nodata  \\
  &  178  &  160  &  12  &  12504  &   141  &  3.92  &  0.04  &  4.05  &   3.5  &   0.1  &   2.9  &   0.3  &  390  &  \nodata  \\
  &  197  &  198  &  15  &  12402  &   182  &  4.15  &  0.06  &  4.25  &   3.2  &   0.1  &   2.2  &   0.5  &  427  &  SB? (Paper 3) \\

\\
NGC4755
  &    1  &  164  &  50  &  15150  &   150  &  4.09  &  0.02  &  4.20  &   4.3  &   0.1  &   2.7  &   0.8  &  447  &  \nodata  \\
  &   15  &   56  &   5  &  21450  &   350  &  3.87  &  0.03  &  3.89  &   8.6  &   0.2  &   5.5  &   1.6  &  446  &  \nodata  \\
  &   25  &  256  &  16  &  20917  &   900  &  3.32  &  0.08  &  3.32  &  11.9  &   1.1  &  12.5  &   1.8  &  348  &  \nodata  \\
  &   54  &  277  &  38  &  10613  &    85  &  3.74  &  0.04  &  4.02  &   2.8  &   0.0  &   2.7  &   0.1  &  363  &  \nodata  \\
  &   71  &  267  &  22  &  21880  &  1650  &  4.42  &  0.17  &  4.52  &   6.8  &   0.9  &   2.4  &   1.7  &  602  &  \nodata  \\
  &   75  &  130  &  14  &  16844  &  1550  &  4.25  &  0.20  &  4.32  &   4.8  &   0.7  &   2.5  &   0.9  &  493  &  \nodata  \\
  &  120  &  183  &  36  &  17495  &  1400  &  4.08  &  0.17  &  4.21  &   5.4  &   0.7  &   3.0  &   0.8  &  476  &  \nodata  \\
  &  144  &  173  &  14  &  10643  &   193  &  3.95  &  0.09  &  4.08  &   2.8  &   0.1  &   2.5  &   0.2  &  373  &  \nodata  \\
  &  172  &   57  &  15  &  23900  &   400  &  3.68  &  0.02  &  3.70  &  11.8  &   0.3  &   8.0  &   1.9  &  432  &  \nodata  \\
  &  195  &   69  &  11  &  27727  &  1350  &  3.61  &  0.12  &  3.63  &  17.5  &   2.0  &  10.6  &   3.3  &  458  &  \nodata  \\
  &  222  &  293  &  23  &  11492  &   232  &  3.90  &  0.08  &  4.18  &   2.9  &   0.1  &   2.3  &   0.3  &  401  &  \nodata  \\
\\
NGC6231
  &   15  &   69  &  10  &  17266  &   200  &  4.32  &  0.02  &  4.35  &   4.9  &   0.1  &   2.5  &   0.3  &  503 &  \nodata  \\
  &   32  &   90  &   9  &  24600  &   400  &  3.77  &  0.03  &  3.82  &  11.6  &   0.3  &   7.0  &   2.3  &  459 &  SB1? \citep{garcia2001}  \\
  &   45  &  176  &  10  &  18100  &   200  &  4.08  &  0.02  &  4.20  &   5.8  &   0.1  &   3.2  &   1.3  &  481 &  \nodata  \\
  &   55  &  311  &  16  &  18250  &    50  &  4.05  &  0.03  &  4.27  &   5.6  &   0.0  &   2.9  &   0.1  &  500 &  \nodata  \\
  &   57  &  184  &  27  &  11119  &   212  &  3.90  &  0.08  &  4.04  &   3.0  &   0.1  &   2.7  &   0.2  &  372 &  \nodata  \\
  &   63  &  119  &  22  &  15450  &    50  &  4.30  &  0.02  &  4.36  &   4.1  &   0.0  &   2.2  &   0.3  &  486 &  \nodata  \\
  &   67  &  266  &  15  &  17733  &   200  &  4.23  &  0.03  &  4.39  &   5.1  &   0.1  &   2.4  &   0.4  &  518  & SB (WEBDA) \\
  &   86  &  176  &  12  &  11703  &   140  &  3.94  &  0.04  &  4.07  &   3.2  &   0.1  &   2.7  &   0.2  &  385 &  \nodata  \\
  &   91  &  137  &  18  &  20700  &   250  &  3.94  &  0.02  &  4.03  &   7.7  &   0.2  &   4.4  &   1.6  &  469  &  SB (WEBDA)  \\
  &  110  &   86  &  11  &  24690  &   400  &  4.03  &  0.03  &  4.06  &  10.5  &   0.3  &   5.0  &   2.7  &  515 &  SB2? \citep{garcia2001}  \\
  &  111  &  402  &  33  &  24000  &   100  &  3.80  &  0.01  &  4.14  &   9.5  &   0.1  &   4.4  &   2.6  &  527 &  SB2 \citep{garcia2001}  \\
  &  114  &  213  &  12  &  15366  &   200  &  4.02  &  0.02  &  4.18  &   4.5  &   0.1  &   2.8  &   0.8  &  446 &  \nodata  \\
  &  116  &   92  &  13  &  26550  &   850  &  4.23  &  0.08  &  4.25  &  10.9  &   0.6  &   4.1  &   0.5  &  581 &  SB1 \citep{garcia2001}  \\
  &  121  &  295  &  31  &  13306  &   304  &  4.04  &  0.07  &  4.29  &   3.5  &   0.1  &   2.2  &   0.2  &  446 &  \nodata  \\
  &  124  &  147  &   9  &  23850  &   150  &  3.90  &  0.02  &  3.98  &  10.2  &   0.1  &   5.4  &   2.4  &  490 &  SB1?  \citep{garcia2001}   \\
  &  132  &  164  &  14  &  28150  &   900  &  4.03  &  0.08  &  4.09  &  13.2  &   0.9  &   5.4  &   3.9  &  556 &  SB2?  (this work)  \\
  &  137  &  257  &  26  &  29194  &     0  &  3.92  &  0.10  &  4.04  &  14.4  &   0.0  &   6.0  &   4.2  &  553 &  SB2?  (this work) \\
  &  156  &  183  &  19  &  14100  &   250  &  4.40  &  0.01  &  4.46  &   3.5  &   0.1  &   1.8  &   0.5  &  494 &  \nodata  \\
  &  162  &   85  &  13  &  28800  &   250  &  3.67  &  0.02  &  3.70  &  18.0  &   0.4  &   9.9  &   4.0  &  480 &  \nodata  \\
  &  165  &  145  &  25  &  \nodata  &  \nodata  &  \nodata  &  \nodata  &  \nodata  &   \nodata  &   \nodata  &   \nodata  &   \nodata  &  \nodata  & SB2 \citep{sana2008}  \\
  &  168  &   62  &   9  &  29472  &   550  &  4.15  &  0.05  &  4.16  &  14.0  &   0.5  &   5.2  &   4.4  &  586 &  \nodata  \\
  &  169  &  261  &  10  &  15350  &    50  &  3.80  &  0.00  &  4.05  &   4.7  &   0.0  &   3.4  &   0.7  &  420 &  \nodata  \\
  &  178  &  165  &  13  &  15633  &   350  &  4.34  &  0.06  &  4.41  &   4.1  &   0.2  &   2.1  &   0.6  &  501 &  \nodata  \\
  &  185  &  299  &  24  &  11655  &   185  &  3.72  &  0.06  &  4.03  &   3.2  &   0.1  &   2.9  &   0.2  &  377 &  \nodata  \\
  &  190  &  178  &   9  &  28850  &   150  &  3.60  &  0.00  &  3.73  &  17.7  &   0.2  &   9.6  &   4.1  &  485 &  \nodata  \\
  &  213  &  324  &  38  &  16966  &   200  &  4.12  &  0.03  &  4.35  &   4.8  &   0.1  &   2.4  &   0.3  &  501 &  \nodata  \\
  &  218  &   87  &   8  &  29600  &   100  &  4.25  &  0.02  &  4.26  &  13.5  &   0.1  &   4.5  &   0.1  &  616 &  SB1? \citep{garcia2001}   \\
  &  235  &  292  &  62  &  12667  &   154  &  3.88  &  0.05  &  4.16  &   3.4  &   0.1  &   2.6  &   0.4  &  412 &  \nodata \\
  &  245  &  350  &  12  &  17900  &     0  &  4.10  &  0.02  &  4.35  &   5.3  &   0.0  &   2.5  &   0.3  &  513 &  \nodata \\
  &  268  &  150  &  21  &  13588  &   239  &  4.13  &  0.05  &  4.22  &   3.7  &   0.1  &   2.4  &   0.6  &  436 &  \nodata \\
\\
\hline
Be Stars: \\
\hline
NGC3293
  &   42  &  259  &  17  &  16205  &  1400  &  4.32  &  0.23  &  4.32  &   4.6  &   0.1  &   2.4  &   0.5  &  494 &  \nodata  \\
  &   46  &  281  &   9  &  29655  &  1400  &  3.88  &  0.20  &  3.88  &  16.9  &   0.6  &   7.8  &   1.6  &  524 &  \nodata  \\
  &   74  &  342  &   6  &  21338  &  1400  &  3.94  &  0.21  &  3.94  &   8.4  &   0.1  &   5.2  &   1.1  &  453 &  \nodata  \\
  &   83  &  365  &  21  &  21778  &  1400  &  3.84  &  0.21  &  3.84  &   9.0  &   0.1  &   5.9  &   1.2  &  440 &  \nodata  \\
  &   99  &  273  &  27  &  29586  &  1400  &  4.12  &  0.20  &  4.12  &  14.3  &   0.5  &   5.4  &   1.1  &  577 &  \nodata  \\
\\
NGC3766 
  &   25  &  261  &   6  &  18995  &   301  &  4.02  &  0.09  &  4.02  &   6.6  &   0.1  &   4.2  &   1.2  &  450  &  \nodata  \\
  &   31  &  197  &   5  &  17834  &   400  &  3.82  &  0.10  &  3.99  &   6.2  &   0.2  &   4.2  &   1.0  &  435  &  \nodata  \\
  &   47  &  190  &   4  &  18399  &   301  &  3.30  &  0.09  &  3.30  &   9.3  &   0.4  &  11.3  &   2.5  &  323  &  \nodata  \\
  &   73  &  296  &   4  &  18274  &   301  &  3.49  &  0.09  &  3.49  &   8.2  &   0.2  &   8.5  &   1.3  &  349  &  \nodata  \\
  &   83  &  111  &  17  &  18817  &   301  &  3.31  &  0.09  &  3.31  &   9.8  &   0.4  &  11.5  &   2.4  &  329  &  \nodata  \\
  &   92  &  214  &  17  &  18725  &   301  &  3.34  &  0.09  &  3.34  &   9.5  &   0.4  &  10.9  &   2.2  &  332  &  \nodata  \\
  &   98  &  265  &  25  &  16890  &   301  &  3.84  &  0.10  &  3.84  &   6.2  &   0.2  &   4.9  &   0.6  &  398  &  \nodata  \\
  &  119  &  215  &  22  &  17792  &   301  &  3.75  &  0.09  &  3.75  &   6.8  &   0.1  &   5.8  &   0.5  &  387  &  \nodata  \\
  &  127  &  347  &   4  &  17687  &   550  &  3.61  &  0.12  &  4.01  &   6.1  &   0.3  &   4.0  &   1.0  &  437  &  \nodata  \\
  &  130  &  285  &   7  &  17519  &   301  &  3.69  &  0.09  &  3.69  &   6.9  &   0.1  &   6.2  &   0.4  &  375  &  \nodata  \\
  &  133  &  215  &  20  &  18564  &   301  &  3.53  &  0.09  &  3.53  &   8.3  &   0.2  &   8.2  &   1.0  &  358  &  \nodata  \\
  &  139  &  343  &  12  &  15945  &   301  &  3.95  &  0.10  &  3.95  &   5.2  &   0.2  &   4.0  &   0.5  &  406  &  \nodata  \\
  &  154  &  112  &  29  &  13254  &  1254  &  3.29  &  0.18  &  3.29  &   5.5  &   0.8  &   8.8  &   3.9  &  282  &  \nodata  \\
  &  196  &  165  &   4  &  19660  &   250  &  3.73  &  0.02  &  3.87  &   7.6  &   0.2  &   5.3  &   1.1  &  426  &  \nodata  \\
  &  198  &  251  &   6  &  19580  &   301  &  4.00  &  0.09  &  4.00  &   7.0  &   0.2  &   4.4  &   1.2  &  449  &  \nodata  \\
  &  200  &  236  &  12  &  16301  &   301  &  3.51  &  0.10  &  3.51  &   6.8  &   0.2  &   7.6  &   1.5  &  337  &  \nodata  \\  
\\
NGC4755
  &   31  &  321  &  34  &  20584  &  1400  &  4.17  &  0.19  &  4.17  &   7.1  &   0.1  &   3.6  &   0.6  &  501 &  \nodata  \\
  &  129  &  238  &  24  &  25040  &  1400  &  3.91  &  0.17  &  3.91  &  11.5  &   0.2  &   6.2  &   1.1  &  483 &  \nodata  \\
  &  171  &  327  &  12  &  26246  &  1400  &  4.04  &  0.17  &  4.04  &  11.6  &   0.2  &   5.4  &   0.9  &  523 &  \nodata  \\
  &  187 & \nodata & \nodata & 16070  & 1400  &  4.23  &  0.21  &  4.23  &  4.7  &  0.1  &  2.8  &   0.5  &  462  &  \nodata  \\ 
\\
NGC6231
  &    2  &  325  &  23  &  15050  &    50  &  3.65  &  0.03  &  4.03  &   4.6  &   0.0  &   3.5  &   0.6  &  412 &  \nodata  \\
  &   79  &  302  &  32  &  16533  &  1400  &  4.18  &  0.17  &  4.18  &   4.9  &   0.1  &   3.0  &   0.4  &  456 &  \nodata  \\
  &   90  &  201  &  25  &  27117  &  1400  &  4.13  &  0.14  &  4.13  &  11.9  &   0.4  &   4.9  &   0.6  &  554 &  SB2? \citep{garcia2001}  \\
  &  221  &  162  &  13  &  14704  &  1400  &  4.19  &  0.19  &  4.19  &   4.1  &   0.1  &   2.7  &   0.3  &  439 &  \nodata  \\
\enddata
\end{deluxetable}

\clearpage

\clearpage

\clearpage

\clearpage

\clearpage

\begin{thebibliography}

\bibitem[Abt et al.(2002)]{abt2002} 
Abt, H.~A., Levato, H., \& Grosso, M.\ 2002, \apj, 573, 359

\bibitem[Balona(1984)]{balona1984} 
Balona, L.~A.\ 1984, \mnras, 211, 973

\bibitem[Carciofi et al.(2006)]{carciofi2006} 
Carciofi, A.~C., et al.\ 2006, \apj, 652, 1617 

\bibitem[Dufton et al.(2006)]{dufton2006} 
Dufton, P.~L., et al.\ 2006, \aap, 457, 265 

\bibitem[Ekstr{\"o}m et al.(2008)]{ekstrom2008} 
Ekstr{\"o}m, S., Meynet, G., Maeder, A., \& Barblan, F.\ 2008, \aap, 478, 467 

\bibitem[Garc{\'{\i}}a \& Mermilliod(2001)]{garcia2001} 
Garc{\'{\i}}a, B., \& Mermilliod, J.~C.\ 2001, \aap, 368, 122

\bibitem[Grundstrom \& Gies(2006)]{grundstrom2006} 
Grundstrom, E.~D., \& Gies, D.~R.\ 2006, \apjl, 651, L53

\bibitem[Heger \& Langer(2000)]{heger2000} 
Heger, A., \& Langer, N.\ 2000, \apj, 544, 1016


\bibitem[Huang \& Gies(2006)]{huang2006b} 
Huang, W., \& Gies, D.~R.\ 2006, \apj, 648, 591 

\bibitem[Huang \& Gies(2008)]{huang2008} 
Huang, W., \& Gies, D.~R.\ 2008, \apj, 683, 1045


\bibitem[Hunter et al.(2008)]{hunter2008} 
Hunter, I., Lennon, D.~J., Dufton, P.~L., Trundle, C., Sim{\'o}n-D{\'{\i}}az, S., Smartt, S.~J., Ryans, R.~S.~I., \& Evans, C.~J. 2008, \aap, 479, 541


\bibitem[Kurucz(1994)]{kurucz1994}
Kurucz, R.\ L. 1994, Kurucz CD-ROM 19, Solar Abundance Model Atmospheres for 0, 1, 2, 4, 8 km/s (Cambridge: SAO)

\bibitem[Lanz \& Hubeny(2003)]{lanz2003}
Lanz, T., \& Hubeny, I. 2003, \apjs, 146, 417

\bibitem[Lanz \& Hubeny(2007)]{lanz2007}
Lanz, T., \& Hubeny, I.\ 2007, \apjs, 169, 83


\bibitem[Martayan et al.(2007)]{martayan2007} 
Martayan, C., Floquet, M., Hubert, A.~M., Guti{\'e}rrez-Soto, J., Fabregat, J., Neiner, C., \& Mekkas, M. 2007, \aap, 472, 577

\bibitem[Martayan et al.(2006)]{martayan2006}
Martayan, C., Fr{\'e}mat, Y., Hubert, A.-M., Floquet, M., Zorec, J., \& Neiner, C.\ 2006, \aap, 452, 273

\bibitem[Mathew et al.(2008)]{mathew2008} 
Mathew, B., Subramaniam, A., \& Bhatt, B.\ C. 2008, \mnras, 388, 1879

\bibitem[McSwain(2008)]{mcswain2008b}
McSwain, M.\ V.  2008, \apj, 686, 1269

\bibitem[McSwain \& Gies(2005a)]{mcswain2005a}
McSwain, M.\ V., \& Gies, D.\ R. 2005a, ApJ, 622, 1052 (Paper 1)

\bibitem[McSwain \& Gies(2005b)]{mcswain2005b}
McSwain, M.\ V., \& Gies, D.\ R. 2005b, ApJS, 161, 118 (Paper 2)

\bibitem[McSwain et al.(2008)]{mcswain2008a} 
McSwain, M.\ V., Huang, W., Gies, D.\ R., Grundstrom, E.\ D., \& Townsend, R.\ H.\ D. 2008, \apj, 672, 590 (Paper 3)

\bibitem[Meilland et al.(2006)]{meilland2006} 
Meilland, A., Stee, P., Zorec, J., \& Kanaan, S. 2006, \aap, 455, 953

\bibitem[Meynet \& Maeder(2000)]{meynet2000} 
Meynet, G., \& Maeder, A.\ 2000, \aap, 361, 101

\bibitem[Napiwotzki et al.(1993)]{napiwotzki1993} 
Napiwotzki, R., Schoenberner, D., \& Wenske, V.\ 1993, \aap, 268, 653

\bibitem[Perry et al.(1991)]{perry1991} 
Perry, C.~L., Hill, G., \& Christodoulou, D.~M. 1991, \aaps, 90, 195 

\bibitem[Porter \& Rivinius(2003)]{porter2003}
Porter, J. M., \& Rivinius, T. 2003, \pasp, 115, 1153

\bibitem[Raboud(1996)]{raboud1996} 
Raboud, D. 1996, \aap, 315, 384

\bibitem[Sana et al.(2008)]{sana2008} 
Sana, H., Gosset, E., Naz\'e, Y., Rauw, G., \& Linder, N. 2008, \mnras, 386, 447

\bibitem[Sana et al.(2006)]{sana2006} 
Sana, H., Gosset, E., Rauw, G., Sung, H., \& Vreux, J.-M.\ 2006, \aap, 454, 1047

\bibitem[Schaller et al.(1992)]{schaller1992}
Schaller, G., Schaerer, D., Meynet, G., \& Maeder, A. 1992, \aaps, 96, 269


\bibitem[von Zeipel(1924)]{vonzeipel1924}
von Zeipel, H. 1924, \mnras, 84, 665 

\bibitem[Wisniewski \& Bjorkman(2006)]{wisniewski2006}
Wisniewski, J.\ P., \& Bjorkman, K.\ S.  2006, ApJ, 652, 458


\end{thebibliography}
\end{document}